\documentclass[12pt,draftclsnofoot,onecolumn]{IEEEtran}

\makeatletter
\usepackage{amsthm,amsmath,amssymb,mathtools,bm,etoolbox,float,hyperref}
\pretocmd\@bibitem{\color{black}\csname keycolor#1\endcsname}{}{\fail}
\newcommand\citecolor[1]{\@namedef{keycolor#1}{\color{black}}}
\usepackage{xcolor,cite,etoolbox}
\DeclareMathOperator*{\argmax}{arg\,max}
\newtheorem*{remark}{Remark}
\usepackage{subcaption}
\usepackage{algorithm}
\usepackage{algpseudocode}
\makeatletter
\renewcommand{\Function}[2]{%
  \csname ALG@cmd@\ALG@L @Function\endcsname{#1}{#2}%
  \def\jayden@currentfunction{#1}%
}
\newcommand{\funclabel}[1]{%
  \@bsphack
  \protected@write\@auxout{}{%
    \string\newlabel{#1}{{\jayden@currentfunction}{\thepage}}%
  }%
  \@esphack
}

\usepackage{tikz,float}
  \newlength\fheight
\newlength\fwidth
\usepackage{tkz-euclide}
\def\BibTeX{{\rm B\kern-.05em{\sc i\kern-.025em b}\kern-.08em
    T\kern-.1667em\lower.7ex\hbox{E}\kern-.125emX}}
    \usepackage{pgfplots} 
\usepackage{pgfgantt}
\usepackage{pdflscape}
\pgfplotsset{compat=newest} 
\pgfplotsset{plot coordinates/math parser=false}
\pgfplotsset{every  tick/.style={black,},ylabel style={font=\tiny},xlabel style={font=\tiny},tick label style={font=\tiny},legend style= {font=\scriptsize},
minor x tick num=1,minor y tick num=1,xminorticks=true,yminorticks=true,}

\citecolor{3gpp2}
\citecolor{3gpp3}
\citecolor{3gpp4}
\citecolor{3gpp5}
\citecolor{3gpp6}
\citecolor{va}
\citecolor{anum}
\citecolor{doubly}
\citecolor{adc}
\citecolor{adc27}
\citecolor{adc36}
\citecolor{adc37}
\citecolor{adc38}
\citecolor{adc39}
\citecolor{lte}
\citecolor{decade}
\citecolor{surv}
\citecolor{tassi}
\citecolor{inv}
\citecolor{urban}
\citecolor{prediction}
\citecolor{pos}
\citecolor{spatially}
\citecolor{cross}
\citecolor{alkhateeb}
\citecolor{ayach}
\citecolor{multibeam}
\citecolor{phase} 
\citecolor{estimation}
\citecolor{fpc}
\citecolor{inversion}
\citecolor{surface5g}
\citecolor{surface6g}
\citecolor{vq1}
\citecolor{vq2}
\citecolor{feedback}
\citecolor{heath}
\citecolor{ni}
\citecolor{cfo}

\begin{document}


\title{\textcolor{black}{Hybrid Precoding for mmWave V2X Doubly-Selective Multiuser MIMO Systems}}


\author{Elyes Balti,~\IEEEmembership{Member,~IEEE} 
\thanks{Elyes Balti is with the Wireless Networking and Communications Group, Department of Electrical and Computer Engineering, The University
of Texas at Austin, Austin, TX 78712 USA e-mail: ebalti@utexas.edu.}}


\maketitle

\begin{abstract}
\textcolor{black}{Millimeter wave (mmWave) is a practical solution to provide high data rate for the vehicle-to-everything (V2X) communications. This enables the future autonomous vehicles to exchange big data with the base stations (BSs) such as the velocity and the location to improve the awareness of the advanced driving assistance system (ADAS). In this context, we consider a single-cell multiuser doubly-selective system wherein the BS simultaneously serves multiple vehicles. To accomplish this requirement, the BS is implemented in hybrid architecture to support multiple spatial streams while the vehicles have analog-only structures. In this work, we develop a low-complexity hybrid precoding algorithm wherein the design of the hybrid precoder at the BS and the analog combiner at the vehicles require small training and feedback overhead. We propose a two-stage hybrid precoding algorithm wherein the first stage designs the analog beamformers as in single user scenario while the second stage designs the multiuser digital precoder at the BS. In the second stage, we derive closed-form digital precoders such as Maximum Ratio Transmission (MRT), Zero-Forcing (ZF) and Minimum Mean Square Error (MMSE) as a first variant while we propose iterative digital precoder as a second variant. The design of the digital precoders for the two variants requires the limited feedback sent from the vehicles to BS. We refer to the random vector quantization  (RVQ) and the beamsteering codebooks to quantize the feedbacks for variants I and II, respectively, since the perfect feedback requires long overhead and large training. We evaluate the rate loss incurred by the quantization of the digital and analog codebooks against the perfect channel state information at the transmitter (CSIT). Moreover, we compare the robustness of the proposed algorithm against the exhaustive (joint) beam search which serves as a benchmarking tool. Besides, we investigate the effects of the vehicle mobility on the error performance. In addition, we quantify the robustness of the systems to support different V2X services in terms of outage probability. Furthermore, we evaluate the impacts of the number of served users against the single user TDMA scenario which serves as benchmarking tool. In addition, we investigate the effects of the number of users and the power of analog-to-digital converter (ADC) on the energy efficiency. Finally, we implemented the proposed systems in mmWave and sub-6 GHz settings as well as for partially and fully connected hardwares.}
\end{abstract}


\IEEEpeerreviewmaketitle

\section{Introduction}

\IEEEPARstart{T}{he} increase demand for bandwidth has been growing over the last few decades, largely due to the increased number of subscribers and the number of communication devices. Due to these factors, the users suffer from the spectrum scarcity as most of the available spectrum is totally assigned to the licensed users. In addition, spectrum sharing-sensing systems which consist of primary and secondary users also reach its bottleneck. Due to the massive number of users, the secondary users become unable to take advantages of the spectrum holes left by the primary users. Also, the primary users are not efficiently communicating since the licensed microwave spectrum becomes very limited. Furthermore, advanced Long Term Evolution (LTE) communications systems are in desperate needs for high-speed wireless data rate in order to share big data, high definition (HD) videos, e.g. advanced driving assistance system (ADAS) for vehicular communications \cite{v1,v2,v3}. 

\subsection{Literature}
\textcolor{black}{Autonomous connected vehicles have recently gained enormous attention in academia and industry. Such networks consist of interconnected vehicles and road side units (RSUs) or base stations (BSs) to exchange relevant data related to velocities, traffic flows, road/weather conditions, safety warning and driving assistance, etc. Such information is important to avoid for self-driving systems to avoid traffic congestion and potential accidents with cyclists and pedestrians. The occurrence of these issues might be pronounced in complicated roads geometry such as the intersections, the corners, where the risk of potential accidents is severe. Typically, the onboard automotive sensors provide the required information for the vehicles to improve the safety awareness or the ADAS.}

\begin{figure}[H]
        \centering
        \begin{subfigure}[b]{0.45\textwidth}
\centering
\includegraphics[width=\linewidth]{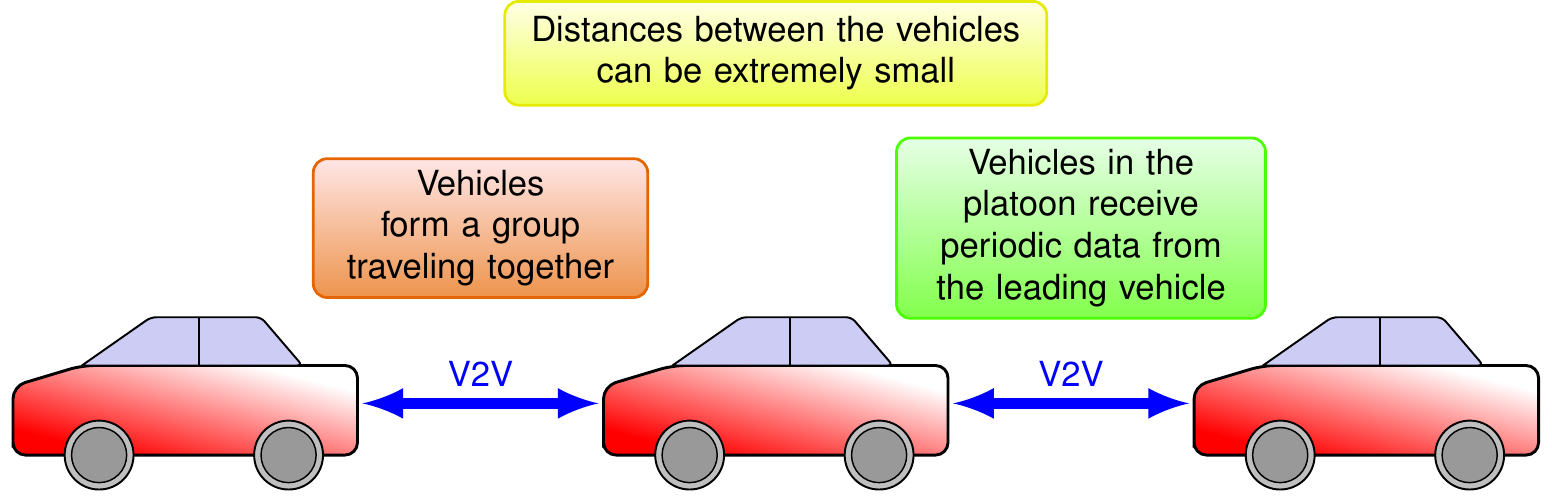}
  \caption{Platooning.}
    \label{a1}   
        \end{subfigure}
        \begin{subfigure}[b]{0.45\textwidth}  
\centering
\includegraphics[width=\linewidth]{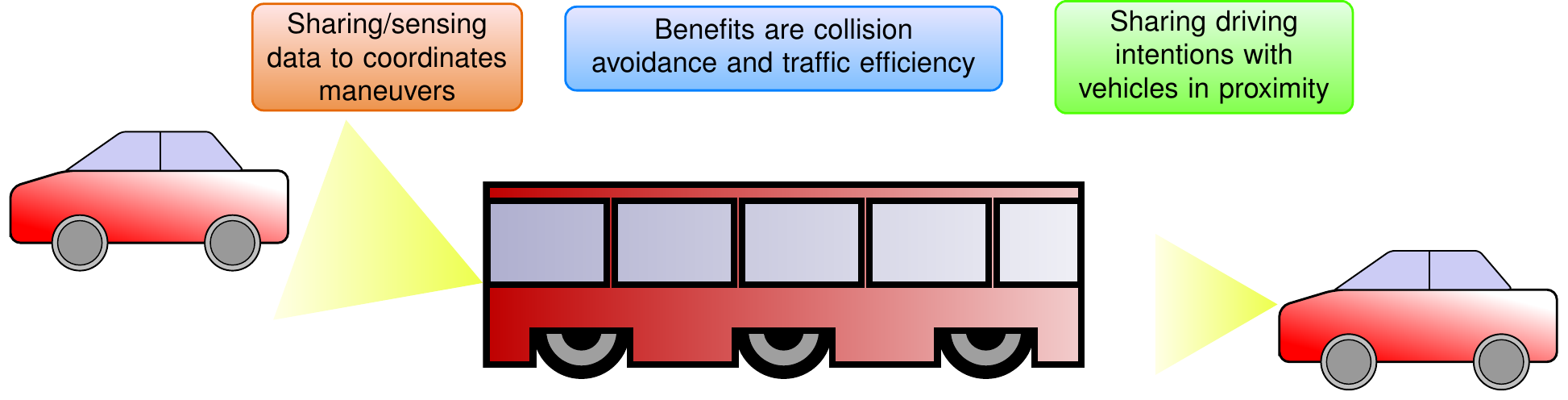}
  \caption{Advanced driving.}
    \label{a2}
        \end{subfigure}
        \vskip\baselineskip
        \vspace*{-.5cm}
        \begin{subfigure}[b]{0.45\textwidth}   
\centering
\includegraphics[width=\linewidth]{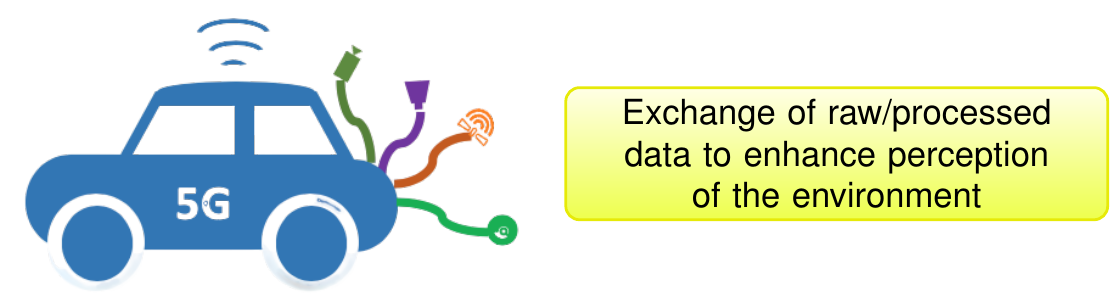}
  \caption{Extended sensors for ADAS enhancement.}
    \label{a3}
        \end{subfigure}
        \begin{subfigure}[b]{0.45\textwidth}   
\centering
\includegraphics[width=\linewidth]{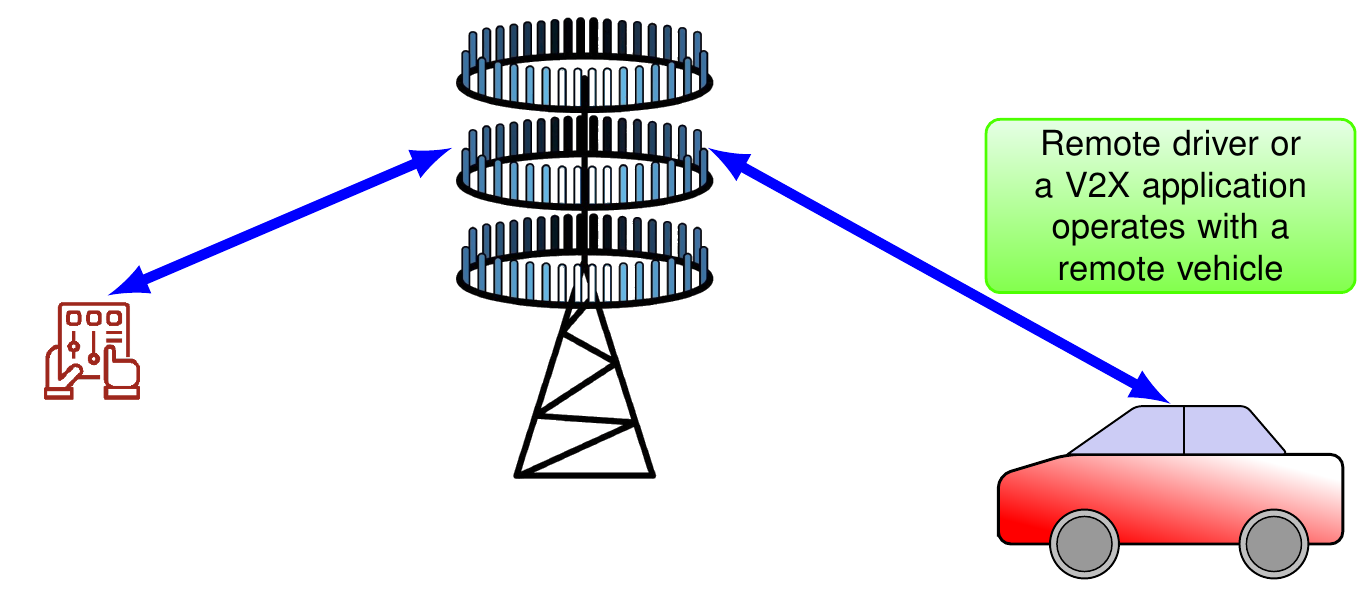}
  \caption{Remote driving/infotainment services.}
    \label{a4}
        \end{subfigure}
        \caption{\textcolor{black}{Detailed use cases into these categories being defined for Release 16.}} 
        \label{services}
    \end{figure}

\textcolor{black}{Connectivity in Vehicular to Everything (V2X) allows the vehicles to engage with future Intelligent Transportation Systems (ITSs) services such as platooning, advanced driving, raw sensor fusion, remote driving and infotainment as illustrated by Fig.~\ref{services} and Table \ref{table2}. Dedicated short-range communication (DSRC) has been implemented for vehicle-to-infrastructure (V2I) and vehicle-to-vehicle (V2V) to exchange basic sensor information within a range of 1 km and with a data rate between 2 to 6 Mbits/s \cite{va}. For LTE cellular system, the maximum data rate is limited to 100 Mbits/s for high mobility while typically much lower rates are achieved \cite{va,lte}. Thereby, conventional DSRC and LTE system cannot support the Gbits/s data rates cannot always meet the communications dictated by \textit{delay and bandwidth sensitive} services, in particular for 5G autonomous vehicles Table \ref{table1}.}

\textcolor{black}{To address these challenges, millimeter wave (mmWave) technology, which refers to the band in the 10 to 300 GHz range\footnote[1]{Although a rigorous definition of mmWave frequencies would place them between 30 and 300 GHz, industry has loosely defined them to include the spectrum from 10 to 300 GHz.}, has been emerged as a promising solution to overcome these shortcoming. In fact, mmWave provides not only a large band of available spectrum, but also a high data rate for exchanging the data. Furthermore, the capacity of the wireless cellular network massively increases and become able to support a large number of subscribers compared to the microwave cellular systems. Hence, mmWave technology is the best way to densify the cellular network and support the technologies that require high data rates \cite{tractable}. However, mmWave suffers from blockage and pathloss since the penetration loss is severe. Designing beamforming (spatial filtering) algorithms for massive antenna arrays can overcome the high propagation losses in mmWave bands, 5G New Radio (NR) can achieve 10x increase in peak and average bit rates over 4G due to the large mmWave bandwidths, e.g. 700 MHz in the 24 GHz band and 850 MHz in the 28 GHz band.  In 2012, prior to the 5G LTE standards, the Wi-Fi IEEE 802.11ad standard adopted usage of the unlicensed 57-64 GHz mmWave band to achieve high bit rates.}

\textcolor{black}{For V2V communications, mmWave has been extensively investigated for more than a decade ago \cite{decade}. Related works have tested mmWave vehicular networks for simpler transceivers and multiple-input-multiple-output (MIMO) large arrays systems \cite{surv,tassi}. In particular, mmWave V2I systems with analog architectures, i.e., the system supports only a single spatial stream, have been considered in urban scenario \cite{urban,inv,pos,prediction} and in the road intersections \cite{spatially,cross} wherein the blockage, coverage, and rate were analyzed. When the V2V sidelink is not available due to the presence of blockage, cooperative communications can solve this shortcoming by offering alternative non-line of sight links to reach the receiver. Cooperative communications can be achieved through relays or intelligent reflecting surfaces. Related work proposed different relaying schemes such as amplify-and-forward for fixed and variable relaying gain, and decode-and-forward relaying, etc \cite{relay,zeroforcing,aggregate}. Although relaying are employed to assist the received signal power and improve the network scalability, they are not power efficient. To address this limitation, passive reflecting surfaces are recently proposed not only to reduce the power consumption but also the energy efficiency (ratio of rate by power consumption), i.e., maximize the rate and minimize the power consumption since the energy efficiency metric becomes the main objective problem for 5g and 6G systems \cite{surface5g,surface6g}. Through cooperative analysis, the rate can be evaluated in the dual-hop system (rate is determined for each hop) and as a result the exchanged rate in V2V sidelink can be estimated.}
\begin{table*}[t]
\renewcommand{\arraystretch}{.7}
\caption{Features of DSRC Vs LTE for V2X \cite{va}.}\label{table1}
    \centering
\textcolor{black}{     \begin{tabular}{|c|c|c|c|}
    \hline
    \textbf{Features} & \textbf{DSRC} & \textbf{D2D LTE-V2X} & \textbf{Cellular LTE-V2X}\\\hline    
    Channel width & 10 MHz & up to 20 MHz & up to 20 MHz  \\\hline
    Frequency band & 5.9 GHz & 5.9 GHz & 450 MHz - 3.8 GHz \\\hline 
    Bit rate & 3-27 Mbps & up to 44 Mbps & up to 75 Mbps  \\\hline
    Range & $\approx$ 100s m & $\approx$ 100s m & up to a few Km  \\\hline
    Spectral efficiency & 0.6 bps/Hz & 0.6 bps/Hz (typical) & 0.6 bps/Hz (typical)  \\\hline
    Coverage & ubiquitous & ubiquitous & ubiquitous  \\\hline
    Mobility support & high speed & high speed & high speed  \\\hline
    Latency & $\times$ ms & $\times$10$\times$100 ms & $\times$10 ms  \\\hline
    \end{tabular}}
\end{table*}

\begin{table*}[t]
\renewcommand{\arraystretch}{.7}
    \caption{Some use cases and network requirements for 5G V2X \cite{3gpp2,3gpp3,3gpp4,3gpp5,3gpp6}}\label{table2}
    \centering
    \textcolor{black}{\begin{tabular}{|c|c|c|c|}
    \hline
    \textbf{Use case} & \textbf{Latency} & \textbf{Reliability} & \textbf{Data rate}\\\hline    
    Vehicle platooning & $\leq$ 25 ms & $\geq$ 90 $\%$ & low  \\\hline
    Remote driving & $\leq$ 5 ms & $\geq$ 99.99 $\%$ & $\geq$ 10 Mbps DL, $\geq$ 20 Mbps UL \\\hline 
    Collective perception of environment & $\leq$ 3 ms & $\geq$ 99 $\%$ & 1 Gbps for a single UE \\\hline
    Cooperative collision avoidance & $\leq$ 10 ms & $\geq$ 99.99 $\%$ & $\geq$ 10 Mbps \\\hline 
    Info sharing for level 2/3 aut. & $\leq$ 100 ms & $\geq$ 90 $\%$ & $\geq$ 0.5 Mbps \\\hline 
    Info sharing for level 4/5 aut. & $\leq$ 100 ms & $\geq$ 90 $\%$ & $\geq$ 50 Mbps \\\hline 
    Video data sharing for improved automated driving & $\leq$ 10 ms & $\geq$ 99.99 $\%$ & $\geq$ 100 - 700 Mbps \\\hline 
    \end{tabular}}
\end{table*}
\textcolor{black}{Unlike single-user MIMO (SU-MIMO) in V2X scenario, considering multiuser MIMO (MU-MIMO) scenario in mmWave systems is challenging wherein the BSs consist of hybrid architecture. Given the hardware constraints as mmWave components are power hungry, the design of analog and digital (hybrid) precoders is complicated in particular for wideband multicarrier systems. In fact, MU-MIMO precoding algorithms have been scanned by the literature wherein the trade-offs come to play between complexity and performance. Due to the sparsity nature of mmWave channels, related work \cite{ayach} exploits this sparsity to develop low-complexity precoding algorithms using the concept of basis pursuit with channel state information at the transmitter (CSIT). In \cite{multibeam}, sequential hybrid precoding has been proposed for SU-MIMO with single spatial stream MIMO-OFDM system wherein the objective problem is based on either the received signal strength (RSS) or the sum rate. In multiuser systems, the baseband precoding layer allows for more degree of freedom (DoF) to reduce the multiuser interference. Therefore, designing low-complexity hybrid precoding algorithms for MU-MIMO mmWave systems is interesting. Besides, related works \cite{phase} proposed the joint analog-digital precoding to achieve both diversity and multiplexing gain. In \cite{estimation}, hybrid precoding algorithms were designed to minimize the the received signal-to-interference-plus-noise ratio (SINR) with quantized phase shifters. Furthermore, \cite{alkhateeb} suggested a two-stage low-complexity hybrid precoding for MU-MIMO narrowband systems with limited feedback wherein the Zero-Forcing (ZF) digital precoder is constructed using the quantized effective channels sent back by each UE to the BS.} 
\subsection{Contributions}
\textcolor{black}{In this paper, we consider a BS serving multiple vehicles simultaneously. Due to the mobility of vehicles, the system is subject to time selectivity which generate the Doppler spread and hence creating the carrier frequency offset (CFO). Moreover, the channels are also vulnerable to frequency selective because of its mmWave natures. Due to time and frequency selectivity, the system is named doubly-selective. In this context, we propose two variants of low-complexity precoding algorithms for MU-MIMO mmWave systems. Each vehicle consists of analog-only architecture and the BS performs in hybrid digital/analog precoding where the number of RF chains is at least as large as the number of vehicles. The proposed algorithms rely on limited feedback to reduce the overhead and yet achieve satisfied performance. The main contributions of this work are summarized in the sequel:
\begin{itemize}
    \item Unlike the work \cite{alkhateeb}, the proposed variant I considers wideband systems with closed-form precoders such as Maximum Ratio Transmission (MRT), ZF and Minimum Mean Square Error (MMSE). The effective subcarriers are quantized using the Random Vector Quantization (RVQ) codebook .
    \item Unlike the work \cite{multibeam}, the proposed variant II generalized the algorithm to support MU-MIMO scenario. The digital precoders are selected from the beamsteering codebook to maximize the sum rate.
    \item Unlike the majority of work related to MU-MIMO \cite{alkhateeb,multibeam}, since we are considering V2X networks,
    we introduce the mobility into the model and we investigate its effects on the system performance.
    \item Consider the exhaustive beam search algorithm which serve as a benchmarking tool for the two proposed variants.
    \item Analyze the system performance in terms of spectral efficiency, probability of error, outage probability and energy efficiency. 
\end{itemize}}

\subsection{Structure}
\textcolor{black}{The rest of the paper is organized as follows. Section II provides a description of channel and system models. In Section III, we formulate the optimization problems and we provide the two variants of MU-MIMO precoding algorithms. In Section IV, we provide the analysis of the reliability metrics in terms of sum rate, outage probability, probability of error and energy efficiency. Numerical results are discussed in Section V while the concluding remarks are summarized in Section VI.}

\subsection{Notation}
\textcolor{black}{We use the following notation throughout the paper. We denote $(\cdot)^T$, $(\cdot)^*$ and $\mathbb{P}[\cdot]$ are the Transpose, Hermitian and the probability measure operators, respectively. Moreover, $\|\cdot\|_F$ and $\det(\cdot)$ are the Frobenius norm and the determinant, respectively.}
\section{System Model}
\begin{figure}[H]
\begin{subfigure}[b]{0.5\textwidth}
\centering
\includegraphics[scale=.5]{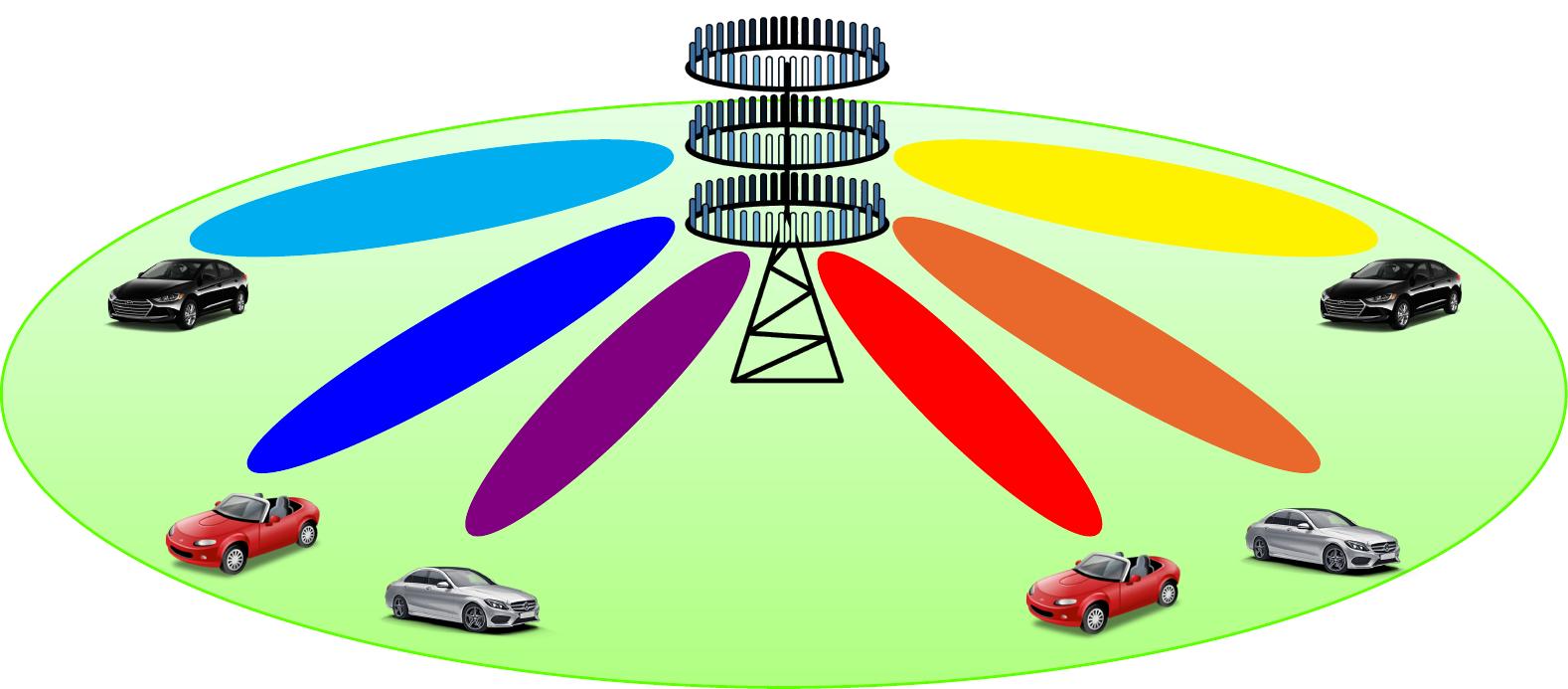}
    \caption{}
    \label{2a}
    \end{subfigure}
    \begin{subfigure}[b]{0.5\textwidth}
\centering
\includegraphics[scale=.55]{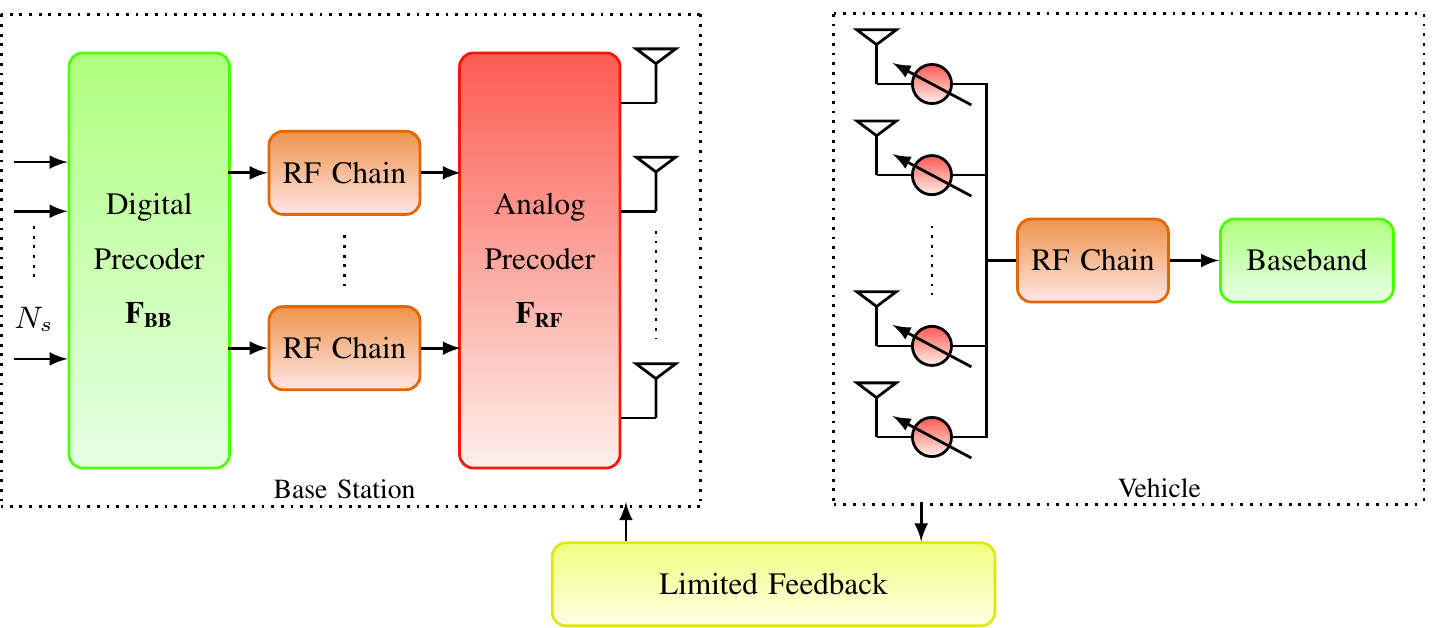}
    \caption{}
    \label{2b}
    \end{subfigure}
    \caption{\textcolor{black}{(a) Illustration of the downlink transmission in a multiuser MIMO systems, where the BS is equipped with $N_{\text{BS}}$ antennas and serves $U$ user terminals simultaneously. This illustration focuses on line-of-sight propagation where the downlink signal can be viewed as angular beams. (b) Physical architectures of the base station and the vehicle. The base station has a hybrid digital/analog architecture supporting multiple spatial streams while the vehicle has analog-only architecture. The vehicle uses its channel estimate to pick the optimal transmitter-side precoder from a codebook known at the BS and the vehicle. For a codebook of size $2^B$, the $B$-bit binary label of the chosen precoder is sent over the feedback. Note that the SNR and or the rate must be known as side information to facilitate communication and is often fed back \cite{feedback}.}}
    \label{system}
\end{figure}

\textcolor{black}{We consider a single-cell multiuser systems illustrated by Fig.~\ref{system}. The BS is equipped with $N_{\text{BS}}$ antennas and $N_{\text{RF}}$ RF chains serves $U$ vehicles simultaneously. Each user equipment (UE) or vehicle is equipped with $N_{\text{UE}}$ antennas and single RF chain as depicted by Fig.~\ref{2b}.}
\subsection{Signal Model}
\textcolor{black}{Each user receives a single spatial stream from the BS and hence we assume that the total number of allowable spatial streams is $N_s = U$. Besides, we assume that the maximum number of served users is equal to the number of BS RF chains, i.e., $U \leq N_{\text{RF}}$. This is motivated by the spatial multiplexing gain of the proposed MU-MIMO hybrid precoding system, which is restricted by $\min\left(N_{\text{RF}},U  \right)$ for $N_{\text{BS}}>N_{\text{RF}}$. For simplicity, the BS uses $U$ out of the $N_{\text{RF}}$ available RF chains to serve the $U$ users. For the $k$-th subcarrier, the BS will apply a $U \times U$ digital precoder $\textbf{F}_{\text{BB}}[k]=\left[\textbf{f}_1^{\text{BB}}[k],\ldots,\textbf{f}_U^{\text{BB}}[k]\right]$ followed by an $N_{\text{BS}}\times U$ analog precoder $\textbf{F}_{\text{RF}}=\left[ \textbf{f}_1^{\text{RF}},\ldots,\textbf{f}_U^{\text{RF}} \right]$. Since the RF precoder is implemented using analog phase-shifters, it has the constant amplitude constraint, i.e., $| [\bold{F}_{\text{RF}}]_{m,n} |^2 = \frac{1}{N_{\text{BS}}}$. Further, we assume that the angles of the analog phase shifters are quantized and have a finite set of possible values. With these assumptions, $\left[\textbf{F}_{\text{RF}}\right]_{m,n} = \frac{1}{\sqrt{N_{\text{BS}}}}e^{j\theta_{m,n}}$, where $\theta_{m,n}$ is the quantized angle. The total power is constrained by normalizing the digital precoder such that $\|\textbf{F}_{\text{RF}}\textbf{F}_{\text{BB}}[k]\|_F^2 = U, k=0,\ldots,K-1$.}

\textcolor{black}{Assuming that the maximum delay spread of the channel is within the cyclic prefix (CP) duration. the proposed system uses OFDM signaling with $K$ subcarriers. At the $k$-th subcarrier, the symbols $\textbf{s}[k] = \left[ \bold{s}_1[k],\ldots,\bold{s}_U[k]\right]$ are transformed to the time-domain using $K$-point IDFT. The CP of length ($L_c$) is then appended to the time domain samples before applying the RF precoder $\textbf{F}_{\text{RF}}$. The OFDM block is formed by the CP followed by the $K$ time-domain samples. The data symbol follows $\mathbb{E}\left[\bold{s}_u[k] \bold{s}^*_u[k]\right] = \frac{\rho}{KU}\boldsymbol{I}$, where $\rho$ is the total average transmit power for the data i.e., without considering the CP, per OFDM symbol. The received signal at the $u$-th user and at the $k$-th subcarrier after combining is} 
\textcolor{black}{\begin{equation}
    \bold{r}_u[k] = \bold{w}^*_u\bold{H}_u[k]\sum_{n=1}^{U}\textbf{F}_{\text{RF}}\textbf{f}_n^{\text{BB}}[k]\bold{s}_n[k] + \bold{w}_u^*\textbf{n}_u[k] 
\end{equation}}
\textcolor{black}{where $\bold{H}_u[k]$ is the $k$-th subcarrier between the BS and $u$-th vehicle, $\bold{w}_u$ is the analog combiner of the $u$-th vehicle, and $\textbf{n}_u[k]$ is the Additive White Gaussian Noise (AWGN) with variance $\sigma^2$ defined as
\begin{equation}
\sigma^2[\text{dBm}] = -173.8 + 10\log_{10}(\text{BW})    
\end{equation}
where $\text{BW}$ is the bandwidth.
\begin{remark}
Note that the analysis for downlink transmission reported in this work can be applied to uplink but with some modifications. For example, Fractional Power Control $\epsilon \in [0~1]$ as used in LTE systems \cite{fpc} can be used. Other power control modes can be considered such as truncated channel inversion power control model \cite{inversion} to determine the transmit power of each uplink user.
\end{remark}}
\subsection{Channel Model}

\textcolor{black}{In this work, we consider doubly-selective MIMO channels having a delay tap length $L$ in the time domain. The $\ell$-th delay tap of the channel is represented by a $N_{\text{RX}} \times N_{\text{TX}}$ matrix, $\ell = 0,\ldots,L-1$,  which, assuming a geometric clusters and rays based channel model given by \cite{anum,doubly} 
\begin{equation}\label{channel}
\begin{split}
\textbf{H}[\ell] =& \sqrt{\frac{N_{\text{RX}}N_{\text{TX}}}{\gamma}} \sum_{c=0}^{C-1}\sum_{r_c=0}^{R_c-1} \alpha_{r_c} p(\ell T_s - \tau_{c}-\tau_{r_c}) \textbf{a}_{\text{RX}}(\theta_c+\vartheta_{r_c})   \textbf{a}_{\text{TX}}^*(\phi_c+\varphi_{r_c}) e^{j\omega_c\ell}   
\end{split}
\end{equation}
where $T_s$ is the signaling interval, $\tau_c$ is the mean time delay of the cluster, $\theta_c$ and $\phi_c$ are the angle of arrival (AoA) and and the angle of departure (AoD), respectively. Each ray has a relative time delay $\tau_{r_c}$, relative AoA ($\vartheta_{r_c}$) AoD ($\varphi_{r_c}$) shifts, and a complex gain $\alpha_{r_c}$. Note that $\gamma$ is the pathloss and $p(\tau)$ is the pulse shaping Raised Cosine filter evaluated at $\tau$. Besides, $\omega_c = 2\pi f_cv_mT_s\sin\left(\theta_c\right)/c_m$ is the normalized Doppler shift, $f_c$ is the carrier frequency, $v_m$ is the maximum relative velocity and $c_m$ is the speed of light.
In addition, $\textbf{a}_{\text{RX}}(\theta)$ and $\textbf{a}_{\text{TX}}(\phi)$ are the antenna array response vectors of the RX and TX, respectively. The array response vector at the RX is given by
\begin{equation}
\textbf{a}_{\text{RX}}(\theta) = \frac{1}{\sqrt{N_{\text{RX}}}}\left[1,e^{j\frac{2\pi d}{\lambda} \sin(\theta)},\ldots,e^{j\frac{2\pi d}{\lambda}\left(N_{\text{RX}} -1\right)\sin(\theta)}  \right]^{T}    
\end{equation}
The channel at the $k$-th subcarrier is given by
\begin{equation}
\textbf{H}[k] = \sum_{\ell=0}^{L-1} \textbf{H}[\ell] e^{-j\frac{2\pi k}{K}\ell}   
\end{equation}
where $K$ is the number of subcarriers in the wideband transmission.}
\subsection{Codebooks}
\textcolor{black}{Due to the constant amplitude constraints in RF hardware, such as the availability of finite set of quantized angles, the design of analog beamformers is limited by the set of feasible beams. Consequently, the beamformers have to be selected from finite codebooks sizes. In this context, related work have suggested different codebooks structures.
\subsubsection{Random Vector Quantization Codebook}
A general framework of Vector Quantization (VQ) codebook was primarily suggested in \cite{vq1}. The idea is to formulate a distortion function (related to SNR or rate loss) and then iteratively minimize this distortion to reach the local optimal solutions. VQ as well as the Grassmanian codebooks are designed for limited feedback assuming that these codebooks are fixed regardless of channel variations. To cope with this limitation, another technique has been proposed to generate random codebook at each cohence fading block in a way to adapt the beams with the channel variations. This new design is called Random VQ (RVQ) codebook. which was first proposed in \cite{vq2}. This codebook design dictates the generation of $2^B$ codebook beams independently and identically distributed according to the stationary distribution of the optimal unquantized beamforming vector. 
\subsubsection{Beamsteering Codebook}
The beams or the spatial filters are designed for the single-path channels. Therefore, the structure of the beams is similar to the array response vector which can be identified by the angle of arrival or departure. Let $\mathcal{W}$ be the feasible set of the beam codebooks with cardinality $|\mathcal{W}|=N_Q$. The codebook, $\mathcal{W}$ consists of the beams $\textbf{a}\left( \frac{2\pi k_Q}{N_Q}\right)$, for the variable $k_Q=0,1,\ldots,N_Q-1$. This codebook is also termed as DFT codebook parametrized with the angle of beam direction.}

\subsection{Hardware Connections}
\subsubsection{Fully-connected structure}
\textcolor{black}{For this structure, each RF chain is connected to all the phase shifters of the array. Although this structure achieves higher rate as it provides more degree of freedom, it is not energy-efficient since a large amount of power is required for the connection between the RF chains and the phase shifters.}
\subsubsection{Partially-connected structure}
\textcolor{black}{For this structure, each RF chain is connected to a sub-array of antennas of the BS array which reduces the hardware complexity in the RF domain. Although fully-connected structure outperforms the partially-connected structure in terms of achievable rate, the latter structure is well advocated for energy-efficient systems. Note that the analog beamformer has the following structure
\begin{equation}
\textbf{F}_{\text{RF}} =
  \begin{pmatrix}
    \textbf{f}_1 & 0 & \dots & 0 \\
    0 & \textbf{f}_2 & \dots & 0 \\
    \vdots & \vdots & \ddots & \vdots \\
    0 & 0 & \dots & \textbf{f}_{N_{\text{RF}}}
  \end{pmatrix}.
\end{equation}
Each RF chain consists of a precoder $\textbf{f}_n,~n = 1\ldots N_{\text{RF}},$ which is a column vector of size $N_{\text{sub}} \times 1$, and $N_{\text{sub}}$ is the number of antennas of the sub-array.}

\section{Beamforming Design}
\textcolor{black}{The objective of this work is to design robust analog and digital beamformers to maximize the sum rate. Based on the received signal at the $u$-th vehicle which is processed by the combiner $\bold{w}_u$, the rate per user is expressed as
\begin{equation} 
\mathcal{I}_u\left({\scriptsize{\textsf{SNR}}}\right) = \frac{1}{K} \sum_{k=0}^{K-1} \log\left( 1 + \frac{  \frac{  {\scriptsize{\textsf{SNR}}}   }{KU}  \left|\bold{w}^*_u \bold{H}_u[k] \bold{F}_{\text{RF}} \bold{f}^{\text{BB}}_u[k] \right|^2  }{     \frac{  {\scriptsize{\textsf{SNR}}}   }{KU} \sum_{n \neq u} \left|\bold{w}^*_u \bold{H}_u[k] \bold{F}_{\text{RF}} \bold{f}^{\text{BB}}_n[k] \right|^2 +1  }   \right)
\end{equation}
The sum rate is expressed as $\mathcal{I}_{\text{sum}}\left({\scriptsize{\textsf{SNR}}}\right) = \sum_{u=1}^U \mathcal{I}_u\left({\scriptsize{\textsf{SNR}}}\right)$. If the sum rate is considered as the objective function or a metric to be optimized, the precoding design problem is to find $\bold{F}_{\text{RF}}^{\star}$, $\left\{ \bold{F}^{\star}_{\text{BB}}[k] \right\}_{k=0}^{K-1}$ and $\left\{ \bold{w}_u^{\star} \right\}_{u=1}^U$ that solve
\begin{equation}\label{metric1}
\begin{split}
&\left\{\bold{F}_{\text{RF}}^{\star},\left\{ \bold{F}^{\star}_{\text{BB}}[k] \right\}_{k=0}^{K-1},\left\{ \bold{w}_u^{\star} \right\}_{u=1}^U\right\} = \argmax ~ \mathcal{I}_{{\scriptsize{\text{sum}}}}({\scriptsize{\textsf{ SNR}}}) \\&
\text{s.t.}~\left[ \bold{F}_{\text{RF}} \right]_{:,u} \in \mathcal{F}, u=1,\ldots,U,\\&~~~~~
\bold{w}_u \in \mathcal{W},~u=1,\ldots,U,\\&~~~~~\| \bold{F}_{\text{RF}} \bold{F}_{\text{BB}}[k]\|_F^2 = U,~k=1,\ldots,K-1
\end{split}
\end{equation}
where $\mathcal{F}$ and $\mathcal{W}$ are the analog beamsteering codebooks at the BS and the vehicles, respectively. Note that we can also consider another cost function or metric to optimize. We define the SINR per user as
\begin{equation}\label{metric2}
\textsf{SINR}_u = \frac{1}{K} \sum_{k=0}^{K-1}   \frac{  \frac{  {\scriptsize{\textsf{SNR}}}   }{KU}  \left|\bold{w}^*_u \bold{H}_u[k] \bold{F}_{\text{RF}} \bold{f}^{\text{BB}}_u[k] \right|^2  }{     \frac{  {\scriptsize{\textsf{SNR}}}   }{KU} \sum_{n \neq u} \left|\bold{w}^*_u \bold{H}_u[k] \bold{F}_{\text{RF}} \bold{f}^{\text{BB}}_n[k] \right|^2 +1  } 
\end{equation}
The sum SINR is expressed as $\textsf{SINR}_{\text{sum}} = \sum_{u=1}^U\textsf{SINR}_u$.
\begin{remark}
The work \cite{multibeam} considered the Received Signal Strength (RSS) as well as the rate as optimization metrics. However, SU-MIMO scenario has been proposed unlike our work which supports MU-MIMO.  
\end{remark}
\subsection{Limited Feedback}
The link adaptation is required to adapt the channel at the transmitter by selecting the modulation and coding schemes (MCS). This selection is occured when the receiver feds back the BS (transmitter) with channel quality indicator (CQI), number of admissible spatial streams or rank indicator (RI) and precoding matrix indicator (PMI). The availability of the channel state information at the transmitter (CSIT) requires large overhead. Besides, the perfect feedback massively grows with the number of antennas as well as the number of users. Therefore, perfect feedback or CSIT becomes impractical. To address this shortcoming, limited or finite-rate feedback has been proposed to overcome the large overhead for the channel state information (CSI) feedback. A common approach to implement the limited feedback is to construct a codebook of quantized precoders known at the transmitter and the receiver.
\subsubsection{Variant I}
After the analog beam training, each vehicle quantizes its effective channel using the RVQ codebook. Then, each vehicle feds back its quantized effective channel to the BS using a finite-rate feedback. Once the BS collects all the quantized effective channels, it constructs the digital precoder which can be MRT, ZF or MMSE in order to simultaneously serve the vehicles and manage the multiuser interference.
\subsubsection{Variant II}
Once the analog beamformers are designed, the BS and the vehicles perform a digital beam training stage using a codebook set of predefined digital precoders given by
\begin{equation}
\mathcal{B} = \left\{ \textbf{B}_1,\ldots, \textbf{B}_{2^{B_{\text{BB}}}}\right\} 
\end{equation}
For variant II, we consider the beamsteering codebook to search for the digital precoders.
\subsection{Beam Search Complexity}
The digital and analog beamforming designs are performed by joint search in the codebooks. A straightforward approach is to go through the exhaustive beam search. Although this approach provides the optimal beamformers, the computational complexity is too costly as the number of operations grows exponentially with the size of codebooks. For this reason, we propose a suboptimal beam search to reduce the complexity cost and maintain an acceptable rate compared to the exhaustive approach. Next, we provide details and complexity comparison between these two approaches.
\subsubsection{Exhaustive Beam Search}
This approach searches for the optimal precoders and combiners by considering all the combinations from the digital and analog beam codebooks. Although this approach is optimal, it is not recommended as it requires high complexity of order $\mathcal{O}\left( 2^{B_{\text{RF}}^{\text{BS}}} \times 2^{B_{\text{RF}}^{\text{Vehicle}}} \times 2^{B_{\text{BB}}} \right)$.
\subsubsection{Sequential Beam Search}
The beamforming design consists of a sequential search wherein the analog beamformers are selected from the DFT codebooks as in single user RF stage followed by a multiuser digital beamforming stage. This approach aims to reduce the complexity of the beamforming design at the expense of the sum rate optimality. Although, this approach is suboptimal, it offers an acceptable overhead and complexity of order $\mathcal{O}\left( 2^{B_{\text{RF}}^{\text{BS}}} \times 2^{B_{\text{RF}}^{\text{Vehicle}}} + 2^{B_{\text{BB}}} \right)$.
\begin{remark}
Note that the number of served users should be reasonable as the digital codebook size grows exponentially with this number.
\end{remark}}
\begin{algorithm}[H]
 \textcolor{black}{\caption{MU-MIMO Hybrid Precoding}
\label{full-digital-beamforming}
\begin{algorithmic}[1]
\State \textbf{Input:} $\bold{H}_u[k],~k=0,\ldots,K-1,~u=1,\ldots,U$
 \State ~~~~~~~~~$\mathcal{F}$ BS analog codebook,~$|\mathcal{F}|=2^{B_{\text{RF}}^{\text{BS}}}$
 \State ~~~~~~~~~$\mathcal{W}$ Vehicle analog codebook,~$|\mathcal{W}|=2^{B_{\text{RF}}^{\text{Vehicle}}}$
  \State ~~~~~~~~~$\mathcal{H}$ RVQ codebook,~$|\mathcal{H}|=2^{B_{\text{BB}}}$
  \State ~~~~~~~~~$\mathcal{B}$ digital codebook,~$|\mathcal{B}|=2^{B_{\text{BB}}}$
\State \textbf{First Stage:} SU-MIMO analog beamforming design
\State For each vehicle $u,~u=1,\ldots,U$
\State ~~~~~~~~The vehicle $u$ and BS select $\bold{v}_u^{\star}$ and $\bold{g}_u^{\star}$ that solve
\begin{equation*}
 \left(\bold{g}_u^{\star},\bold{v}_u^{\star}\right) =    \argmax\limits_{\textbf{g}_u\in \mathcal{W},~\textbf{v}_u \in \mathcal{F}} \sum_{k=1}^K \left|\bold{g}_u^*\bold{H}_u[k]\bold{v}_u \right|^2
\end{equation*}
\State ~~~~~~~~Vehicle $u$ sets $\bold{w}_u=\bold{g}_u^{\star}$
\State BS sets $\bold{F}_{\text{RF}} = \left[\bold{v}_1^{\star},\bold{v}_2^{\star},\ldots,\bold{v}_U^{\star} \right]$
\State \textbf{Second Stage:} MU-MIMO digital precoding design
\State \textbf{Variant I: Closed-Form Precoder}
\For{$k\gets 0:K-1$}
\State For each vehicle $u,~u=1,\ldots,U$
\State ~~~~~~~~Vehicle $u$ estimates its effective subcarrier $\overline{\bold{h}}_u[k]=\bold{w}_u^*\bold{H}_u[k]\bold{F}_{\text{RF}}$
\State ~~~~~~~~Vehicle $u$ quantizes $\overline{\bold{h}}_u[k]$ using the RVQ codebook and feeds back $\widehat{\bold{h}}_u[k]$ where
\begin{equation*}
\widehat{\bold{h}}_u[k] =    \argmax\limits_{ \bold{h}\in \mathcal{H} } \| \overline{\bold{h}}_u[k] \bold{h}  \|
\end{equation*}
\State For MRT precoding, the BS designs $\bold{F}_{\text{BB}}[k]=\widehat{\bold{H}}^*[k]$,~$\widehat{\bold{H}}[k]=\left[ \widehat{\bold{h}}_1[k],\ldots,\widehat{\bold{h}}_U[k]  \right]^*$
\State For ZF precoding, the BS designs $\bold{F}_{\text{BB}}[k]=\widehat{\bold{H}}^*[k]\left( \widehat{\bold{H}}[k] \widehat{\bold{H}}^*[k]  \right)^{-1}$
\State For MMSE precoding, the BS designs $\bold{F}_{\text{BB}}[k]=\widehat{\bold{H}}^*[k]\left( \widehat{\bold{H}}[k] \widehat{\bold{H}}^*[k] + \frac{KU}{ {\scriptsize{\textsf{SNR}}} }\boldsymbol{I} \right)^{-1}$
\EndFor
\State \textbf{Variant II: Iterative Precoder}
\For{$k\gets 0:K-1$}
\begin{equation}
\textbf{F}^{\star}_{\text{BB}}[k] = \left\{\textbf{f}^{\star \text{BB}}_u[k]\right\}_{u=1}^U = \argmax\limits_{\textbf{f}_u \in \mathcal{B}}~ \mathcal{I}_{{\scriptsize{\text{sum}}}}({\scriptsize{\textsf{ SNR}}})
\end{equation}
\EndFor
\State \Return $\bold{F}^{\star}_{\text{RF}},~\left\{\bold{F}^{\star}_{\text{BB}}[k]\right\}_{k=0}^{K-1},~\left\{\bold{w}^{\star}_u\right\}_{u=1}^U$
\end{algorithmic}}
\end{algorithm}

\section{Performance Analysis}
\subsection{Outage Probability}
\textcolor{black}{
Once a transmission strategy is specified, the corresponding outage probability for rate $R$ (bits/s/Hz) is then
\begin{equation}
    P_{\textsf{out}}(\textsf{\scriptsize{SNR}}, R) = \mathbb{P}[\mathcal{I}(\textsf{\scriptsize{SNR}})<R].
\end{equation}
With convenient powerful channel codes, the probability of error when there is no outage is very small and hence the outage probability is an accurate approximation for the actual block error probability. As justified in the literature, modern radio systems such as UMTS and LTE operate at a target error probability. Therefore, the primary performance metric is the maximum rate\footnote[2]{\textcolor{black}{In this work, we define the notion of rate with outage as the average data rate that is correctly received/decoded at the receiver which is equivalent to the throughput. In other standards in the literature, the rate with outage is assimilated with the transmit data rate. The only difference is if we consider rate with outage as the throughput, we account for the probability of bursts (outage) and we multiply by the term (1-$\epsilon$), while for the transmit data rate, the term (1-$\epsilon$) is not accounted anymore.}}, at each {\textsf{\scriptsize{SNR}}}, such that this threshold is not overtaken, i.e.,
\begin{equation}
R_\epsilon(\textsf{\scriptsize{SNR}}) = \max_{\zeta}\left\{ \zeta: P_{\textsf{out}}(\textsf{\scriptsize{SNR}}, \zeta) \leq \epsilon \right\}
\end{equation}
where $\epsilon$ is the target.}

\subsection{Bit Error Probability}
\textcolor{black}{The work \cite{multibeam} considered the probability of error as a cost function to evaluate their beamforming algorithms. We also consider this performance to measure the resiliency of the proposed variants. The probability of error is also considered as a reliability metric for V2X systems. For example, the mobility introduces the Doppler shift which in turn creates the carrier frequency offset (CFO) that rotates the constellation points. Therefore, a good metric to evaluate the impacts of the CFO is nothing but the probability of error. In this work, we consider the QPSK modulation as well as the low-density parity check (LDPC) as channel coding as standardized by the 5G New Radio (NR).}
\subsection{Energy Efficiency}
\textcolor{black}{The energy efficiency, expressed in bits/s/Hz/Watt or bits/s/Joule, is defined as the ratio between the spectral efficiency and the total power consumption. It is expressed as
\begin{equation}
\mathcal{J}(\rho) = \frac{\mathcal{I}(\rho)}{\rho_{\text{Total}}}    
\end{equation}
where $\rho_{\text{Total}}$ is the total power consumption. For analog, hybrid, and full-digital combiners (AC, DC and HC) wherein the total power consumption model for each architecture is defined by \cite{adc}
\begin{equation}
\rho^{\text{AC}}_{\text{Total}} = N_{\text{RX}}\left( \rho_{\text{LNA}} + \rho_{\text{PS}}\right) + \rho_{\text{RF}} + \rho_{\text{C}} + 2\rho_{\text{ADC}}   
\end{equation}
\begin{equation}
\rho^{\text{DC}}_{\text{Total}} =   N_{\text{RX}} \left(\rho_{\text{LNA}} + \rho_{\text{RF}} + 2\rho_{\text{ADC}}  \right) 
\end{equation}
\begin{equation}
\rho^{\text{HC}}_{\text{Total}} = N_{\text{RX}}\left( \rho_{\text{LNA}} + \rho_{\text{SP}}  \right)+  N_{\text{PS}}\rho_{\text{PS}} +  N_{\text{RF}} \left( \rho_{\text{RF}} + \rho_{\text{C}} + 2\rho_{\text{ADC}} \right)   
\end{equation}
where $\rho_{\text{RF}}$ is the power consumption per RF chain which is defined by
\begin{equation}
\rho_{\text{RF}} = \rho_{\text{M}} + \rho_{\text{LO}} + \rho_{\text{LPF}} + \rho_{\text{B}\text{B}_{\text{amp}} }    
\end{equation}
Note that $N_{\text{PS}}$ is given by
\begin{equation}
N_{\text{PS}} = \left\{
        \begin{array}{ll}
            N_{\text{RX}}N_{\text{RF}} & \quad \text{Fully-connected} \\
            N_{\text{RX}} & \quad \text{Partially-connected}
        \end{array}
    \right. .
\end{equation}}
\begin{table}[H]
\renewcommand{\arraystretch}{.7}
\caption{Power Consumption of Each Device \cite{adc}.}
\label{power}
\centering
\textcolor{black}{
\begin{tabular}{rll}
\bfseries Device & \bfseries Notation & \bfseries Value\\
\hline
Low Noise Amplifier (LNA) \cite{adc36} & $\rho_{\text{LNA}}$ & 39 mW\\
Splitter & $\rho_{\text{SP}}$ & 19.5 mW\\
Combiner \cite{adc36} & $\rho_{\text{C}}$ & 19.5 mW\\
Phase Shifter \cite{adc37,adc38} & $\rho_{\text{PS}}$ & 2 mW\\
Mixer \cite{adc39} & $\rho_{\text{M}}$  & 16.8 mW\\
Local Oscillator \cite{adc27} & $\rho_{\text{LO}}$ & 5 mW\\
Low Pass Filter \cite{adc27} & $\rho_{\text{LPF}}$ & 14 mW\\
Baseband Amplifier \cite{adc27} & $\rho_{\text{B}\text{B}_{\text{amp}} }$ & 5 mW\\
ADC & $\rho_{\text{ADC}}$ & Table \ref{setting}
\end{tabular}}
\end{table}
\section{Numerical Analysis}
In this section, we discuss the numerical results of the reliability metrics performed by Monte Carlo simulations\footnote[3]{For all cases, 1000 realizations of the random variables were generated to perform the Monte Carlo simulation in MATLAB.}. Unless otherwise stated, Table \ref{param} illustrates the values of the system parameters. 
\begin{table}[H]
\renewcommand{\arraystretch}{.7}
\caption{System Parameters \cite{anum}.}
\label{param}
\centering
\textcolor{black}{
\begin{tabular}{rll}
\bfseries Parameter & \bfseries mmWave & \bfseries Sub-6 GHz\\
\hline
Carrier frequency & 28 GHz & 3.5 GHz\\
Bandwidth &850 MHz& 150 MHz\\
Number of antennas at vehicle& 4& 2\\
Number of antennas at BS & 32 & 16 \\
Antenna separation & $\lambda/2$&$\lambda/2$\\
Antenna correlation & None & None\\
Number of clusters & 4 & 10\\
Number of rays per cluster & 10 & 20\\
Angular spread& 2$^{\circ}$ & 2$^{\circ}$\\
Pathloss exponent& 3&3\\
Number of subcarriers ($K$) & 16& 32\\
Cyclic prefix& $K/4$ & $K/4$\\
Number of RF chains & 2 & 2\\
Number of spatial streams & 2 & 2\\
Roll-off factor & 1 & 1\\
Velocity ($v_m$) & 20 km/h & 20 km/h\\ 
Channel coding & LDPC & LDPC\\
Modulation & QPSK & QPSK\\
Coding rate & 1/2 & 1/2
\end{tabular}}
\end{table}
\textcolor{black}{\begin{figure}[H]
\centering
\setlength\fheight{5.5cm}
\setlength\fwidth{7.5cm}
%
%
\definecolor{mycolor1}{rgb}{0.00000,0.44700,0.74100}%
\definecolor{mycolor2}{rgb}{0.85000,0.32500,0.09800}%
\definecolor{mycolor3}{rgb}{0.92900,0.69400,0.12500}%
\definecolor{mycolor4}{rgb}{0.49400,0.18400,0.55600}%
\begin{tikzpicture}

\begin{axis}[%
width=0.951\fwidth,
height=\fheight,
at={(0\fwidth,0\fheight)},
scale only axis,
xmin=-20,
xmax=10,
xlabel style={font=\color{white!15!black}},
xlabel={\textsf{SNR (dB)}},
ymin=0,
ymax=4.5,
ylabel style={font=\color{white!15!black}},
ylabel={\textsf{Average Rate per User (bits/s/Hz)}},
axis background/.style={fill=white},
axis x line*=bottom,
axis y line*=left,
legend style={at={(0.03,0.97)}, anchor=north west, legend cell align=left, align=left, draw=none}
]

\node[right, align=left, rotate=20]
at (axis cs:1,1) {$\scriptsize{ \textsf{B}_{\textsf{BB}} = \textsf{10~bits}    }$};

\node[right, align=left, rotate=42]
at (axis cs:0,2.45) {\scriptsize\sffamily{CSIT}};

\addplot [color=black,dash pattern={on 10pt off 1pt on 0pt off 0pt} ,line width=1.3pt]
  table[row sep=crcr]{%
-21.2339595381851	0.0683737572156085\\
-19.2339595381851	0.153285992532198\\
-17.2339595381851	0.243793854981072\\
-15.2339595381851	0.336283441517683\\
-13.2339595381851	0.434221716359153\\
-11.2339595381851	0.537938778200082\\
-9.2339595381851	0.648563470147258\\
-7.2339595381851	0.766431173051843\\
-5.2339595381851	0.89170406315662\\
-3.2339595381851	1.02414915075868\\
-1.2339595381851	1.16329382704463\\
0.766040461814896	1.30840450336511\\
2.7660404618149	1.45867120619965\\
4.7660404618149	1.61296967546663\\
6.7660404618149	1.77068006046212\\
8.7660404618149	1.92968629774723\\
10.7660404618149	2.09176815830827\\
};
\addlegendentry{\scriptsize{\textsf{MRT}}}

\addplot [color=black,dash pattern={on 10pt off 1pt on 1pt off 1pt} ,line width=1.3pt]
  table[row sep=crcr]{%
-21.2339595381851	0.058155177916501\\
-19.2339595381851	0.143329287104189\\
-17.2339595381851	0.235299492627075\\
-15.2339595381851	0.329729680930489\\
-13.2339595381851	0.430884897129941\\
-11.2339595381851	0.53927915607679\\
-9.2339595381851	0.656413897838032\\
-7.2339595381851	0.782858703772945\\
-5.2339595381851	0.918978234542209\\
-3.2339595381851	1.06464629854644\\
-1.2339595381851	1.21941814912249\\
0.766040461814896	1.38246675482349\\
2.7660404618149	1.55283842581652\\
4.7660404618149	1.7290354141437\\
6.7660404618149	1.91030904471468\\
8.7660404618149	2.09348763742717\\
10.7660404618149	2.28142262834999\\
};
\addlegendentry{\scriptsize{\textsf{ZF}}}

\addplot [color=black, line width=1.3pt]
  table[row sep=crcr]{%
-21.2339595381851	0.0721962051533418\\
-19.2339595381851	0.169497767641213\\
-17.2339595381851	0.273606474326461\\
-15.2339595381851	0.380125440645053\\
-13.2339595381851	0.493271220462233\\
-11.2339595381851	0.613443164786006\\
-9.2339595381851	0.742010741040763\\
-7.2339595381851	0.87937640455167\\
-5.2339595381851	1.02573011442992\\
-3.2339595381851	1.1807801597861\\
-1.2339595381851	1.34394285531699\\
0.766040461814896	1.51431932570769\\
2.7660404618149	1.69091834148098\\
4.7660404618149	1.87237612585565\\
6.7660404618149	2.05793390898502\\
8.7660404618149	2.24505104231171\\
10.7660404618149	2.43584907592121\\
};
\addlegendentry{\scriptsize{\textsf{MMSE}}}

\addplot [color=black, line width=1.3pt]
  table[row sep=crcr]{%
-21.2339595381851	0.0802225714703887\\
-19.2339595381851	0.210446020584028\\
-17.2339595381851	0.354533932845744\\
-15.2339595381851	0.503910044157884\\
-13.2339595381851	0.667557680957663\\
-11.2339595381851	0.847114282224531\\
-9.2339595381851	1.04633825671533\\
-7.2339595381851	1.26726805904917\\
-5.2339595381851	1.51157537872142\\
-3.2339595381851	1.7798898063535\\
-1.2339595381851	2.07206211317433\\
0.766040461814896	2.38680961830844\\
2.7660404618149	2.72241695278323\\
4.7660404618149	3.07512258586053\\
6.7660404618149	3.44348631347501\\
8.7660404618149	3.81760516708137\\
10.7660404618149	4.20730399433681\\
};

\addplot [color=black, line width=.5pt]
  table[row sep=crcr]{%
5.17677669529664	1.57322330470336\\
5.16865369378648	1.5654575074077\\
5.16018990182251	1.55806460630177\\
5.15140242178442	1.55105953986729\\
5.14230901011288	1.54445646290702\\
5.13292804143	1.5382687179428\\
5.12327847141059	1.53250880825498\\
5.11337979847933	1.52718837261762\\
5.10325202441125	1.52231816178057\\
5.09291561391508	1.51790801674598\\
5.08239145328122	1.51396684888302\\
5.07170080817777	1.51050262192113\\
5.06086528068006	1.50752233585805\\
5.04990676562034	1.50503201281531\\
5.03884740634593	1.50303668486961\\
5.02770954997525	1.50154038388469\\
5.01651570224204	1.5005461333644\\
5.00528848201912	1.50005594234322\\
4.99405057561347	1.50007080132672\\
4.98282469092518	1.50059068029004\\
4.97163351156259	1.50161452873864\\
4.96049965100666	1.5031402778309\\
4.94944560691699	1.50516484455861\\
4.93849371567185	1.5076841379766\\
4.92766610723414	1.51069306746913\\
4.91698466043444	1.51418555303626\\
4.90647095876151	1.51815453757943\\
4.89614624674953	1.52259200116137\\
4.88603138705033	1.52748897721163\\
4.87614681827719	1.53283557064488\\
4.86651251370554	1.53862097785545\\
4.85714794091391	1.54483350854766\\
4.84807202244671	1.55146060935782\\
4.83930309757837	1.55848888922026\\
4.83085888525602	1.56590414642594\\
4.82275644829569	1.57369139731924\\
4.81501215890428	1.58183490657469\\
4.80764166559702	1.59031821899255\\
4.80065986157727	1.59912419274911\\
4.79408085464254	1.60823503403431\\
4.78791793867752	1.61763233300686\\
4.78218356679178	1.62729710099411\\
4.77688932615637	1.63720980886157\\
4.77204591459016	1.64735042647451\\
4.76766311894329	1.65769846317183\\
4.76374979532133	1.66823300917064\\
4.76031385119018	1.67893277781763\\
4.75736222939784	1.68977614860195\\
4.75490089414531	1.70074121084282\\
4.75293481893503	1.71180580796331\\
4.75146797652117	1.72294758226116\\
4.75050333088203	1.73414402008586\\
4.75004283123093	1.74537249733091\\
4.7500874080775	1.75661032514932\\
4.75063697134744	1.76783479579981\\
4.75169041056451	1.77902322853131\\
4.75324559709449	1.79015301541284\\
4.75529938844631	1.8012016670163\\
4.757847634622	1.81214685785974\\
4.76088518650238	1.82296647151945\\
4.76440590625163	1.83363864531951\\
4.7684026797197	1.84414181450867\\
4.7728674308176	1.85445475583524\\
4.77779113783627	1.86455663043185\\
4.78316385167641	1.87442702592359\\
4.78897471595209	1.8840459976743\\
4.79521198892775	1.89339410908776\\
4.80186306724417	1.90245247088227\\
4.80891451138544	1.91120277925934\\
4.81635207283562	1.91962735288928\\
4.82416072286994	1.92770916863903\\
4.8323246829227	1.93543189596999\\
4.84082745647023	1.94277992993635\\
4.84965186236464	1.94973842271726\\
4.85878006955094	1.95629331361913\\
4.86819363309737	1.96243135748738\\
4.87787353146627	1.96814015147029\\
4.88780020494988	1.97340816008085\\
4.89795359519382	1.978224738506\\
4.90831318572797	1.98258015411605\\
4.91885804342313	1.98646560613095\\
4.92956686078964	1.98987324340362\\
4.94041799903235	1.99279618028439\\
4.95138953177513	1.99522851053448\\
4.96245928936644	1.99716531926046\\
4.97360490367648	1.99860269284558\\
4.98480385329542	1.99953772685775\\
4.9960335090413	1.99996853191847\\
5.00727117968578	1.99989423752055\\
5.01849415780518	1.99931499378713\\
5.02967976566433	1.99823197116832\\
5.04080540104032	1.99664735807614\\
5.05184858289379	1.99456435646248\\
5.06278699679527	1.99198717534909\\
5.07359854001488	1.9889210223226\\
5.08426136618426	1.98537209301181\\
5.09475392944048	1.98134755856847\\
5.10505502796272	1.97685555117685\\
5.11514384681383	1.97190514762149\\
5.125	1.96650635094611\\
};

\end{axis}
\end{tikzpicture}%
    \caption{\textcolor{black}{Spectral efficiency performance results: The number of served users is 4 users and the BS hardware is implemented in fully-connected structure. We also consider 3 variants of digital precoders such as MRT, ZF and MMSE. The simulations are performed using mmWave configuration.}}
    \label{plot1}
\end{figure}
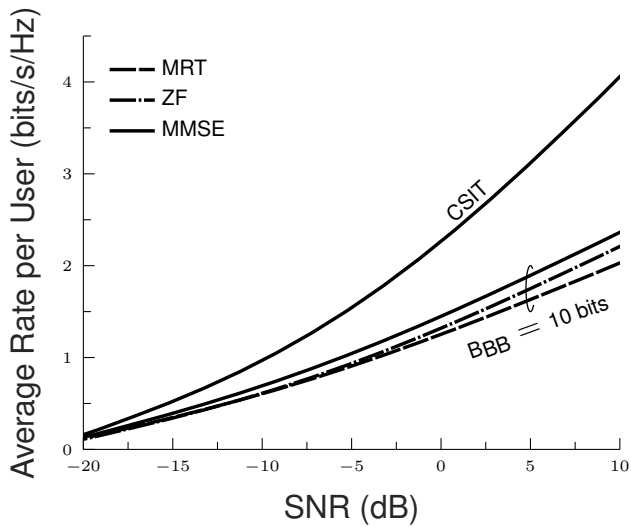
Fig.~\ref{plot1} illustrates the impact of the number of quantization bits on the spectral efficiency. When the effective channel is poorly quantized (10 bits), the loss introduced by the multiuser interference is larger. However, the rate loss per user decreases when the number of quantization bits increases and the achievable rate per user becomes closer to the perfect CSIT scenario. Nonetheless, high level quantization is not practical as it requires long overhead and hefty feedback rate. We further observe that MMSE outperforms ZF and MRT as usual, however, MMSE and ZF performances become identical at high SNR range which cannot be reached at mmWave frequencies.
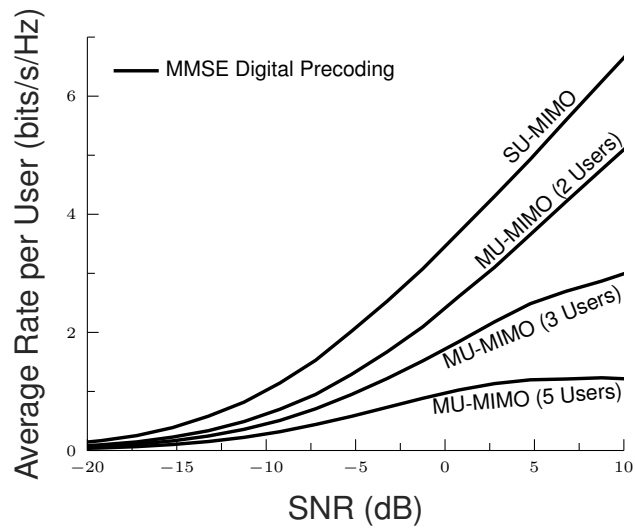
\begin{figure}[H]
\centering
\setlength\fheight{5.5cm}
\setlength\fwidth{7.5cm}
%
%
\definecolor{mycolor1}{rgb}{0.00000,0.44700,0.74100}%
\definecolor{mycolor2}{rgb}{0.85000,0.32500,0.09800}%
\definecolor{mycolor3}{rgb}{0.92900,0.69400,0.12500}%
\definecolor{mycolor4}{rgb}{0.49400,0.18400,0.55600}%
\definecolor{mycolor5}{rgb}{0.46600,0.67400,0.18800}%
\begin{tikzpicture}

\begin{axis}[%
width=0.951\fwidth,
height=\fheight,
at={(0\fwidth,0\fheight)},
scale only axis,
xmin=-20,
xmax=10,
xlabel style={font=\color{white!15!black}},
xlabel={\textsf{SNR (dB)}},
ymin=0,
ymax=7,
ylabel style={font=\color{white!15!black}},
ylabel={\textsf{Average Rate per User (bits/s/Hz)}},
axis background/.style={fill=white},
axis x line*=bottom,
axis y line*=left,
legend style={at={(0.03,0.97)}, anchor=north west, legend cell align=left, align=left, draw=none}
]

\node[right, align=left, rotate=45]
at (axis cs:3,4.8) {\scriptsize\sffamily{SU-MIMO}};

\node[right, align=left, rotate=43]
at (axis cs:1.5,3.0) {\scriptsize\sffamily{MU-MIMO (2 Users)}};

\node[right, align=left, rotate=22]
at (axis cs:-.6,1.4) {\scriptsize\sffamily{MU-MIMO (3 Users)}};

\node[right, align=left, rotate=5]
at (axis cs:-1.3,.7) {\scriptsize\sffamily{MU-MIMO (5 Users)}};

\addplot [color=black, line width=1.3pt]
  table[row sep=crcr]{%
-21.2339595381851	0.107183153484176\\
-19.2339595381851	0.166470598837473\\
-17.2339595381851	0.252643028534521\\
-15.2339595381851	0.387692546170135\\
-13.2339595381851	0.580775920118016\\
-11.2339595381851	0.819436557726156\\
-9.2339595381851	1.14277585273552\\
-7.2339595381851	1.52827997328402\\
-5.2339595381851	2.01795878358508\\
-3.2339595381851	2.52843345672164\\
-1.2339595381851	3.07945937613155\\
0.766040461814896	3.69082457973615\\
2.7660404618149	4.30281023328404\\
4.7660404618149	4.9298692426525\\
6.7660404618149	5.59523366986331\\
8.7660404618149	6.25004559396327\\
10.7660404618149	6.89992777055557\\
};
\addlegendentry{\scriptsize{\textsf{MMSE Digital Precoding}}}

\addplot [color=black, line width=1.3pt]
  table[row sep=crcr]{%
-21.2339595381851	0.0619548570681087\\
-19.2339595381851	0.0965091015328716\\
-17.2339595381851	0.14566835090416\\
-15.2339595381851	0.226934566134519\\
-13.2339595381851	0.335015298075085\\
-11.2339595381851	0.490604774730359\\
-9.2339595381851	0.69834093334218\\
-7.2339595381851	0.949828682918941\\
-5.2339595381851	1.28800158414506\\
-3.2339595381851	1.67099864809288\\
-1.2339595381851	2.09666348871611\\
0.766040461814896	2.60669869697954\\
2.7660404618149	3.10152968210096\\
4.7660404618149	3.65618468106439\\
6.7660404618149	4.21312333214935\\
8.7660404618149	4.76067604792098\\
10.7660404618149	5.30519336115445\\
};

\addplot [color=black, line width=1.3pt]
  table[row sep=crcr]{%
-21.2339595381851	0.0443757380237179\\
-19.2339595381851	0.0696125188069121\\
-17.2339595381851	0.106999916136759\\
-15.2339595381851	0.162873385048217\\
-13.2339595381851	0.245226430846297\\
-11.2339595381851	0.35873503851239\\
-9.2339595381851	0.50861562370568\\
-7.2339595381851	0.706477573267481\\
-5.2339595381851	0.947777362417953\\
-3.2339595381851	1.2191915143368\\
-1.2339595381851	1.51901422055347\\
0.766040461814896	1.84129176446617\\
2.7660404618149	2.17831088557114\\
4.7660404618149	2.4856411640282\\
6.7660404618149	2.69579325453126\\
8.7660404618149	2.86907149119194\\
10.7660404618149	3.07048145323259\\
};


\addplot [color=black, line width=1.3pt]
  table[row sep=crcr]{%
-21.2339595381851	0.0274269341572118\\
-19.2339595381851	0.042755170986422\\
-17.2339595381851	0.0656670653296228\\
-15.2339595381851	0.101166177926236\\
-13.2339595381851	0.151896117833403\\
-11.2339595381851	0.222595581363543\\
-9.2339595381851	0.317919742427573\\
-7.2339595381851	0.440733633332963\\
-5.2339595381851	0.578301134781423\\
-3.2339595381851	0.731371882740628\\
-1.2339595381851	0.884752949905139\\
0.766040461814896	1.02428811303396\\
2.7660404618149	1.13267325178593\\
4.7660404618149	1.19681679889942\\
6.7660404618149	1.21057501718385\\
8.7660404618149	1.2313623308215\\
10.7660404618149	1.20468124809562\\
};

\end{axis}
\end{tikzpicture}%
    \caption{\textcolor{black}{Spectral efficiency performance results: We investigate the effects of the number of users on the average rate. The performance is implemented using MMSE digital precoding and mmWave configuration. Note that the effective subcarriers and the analog codebooks are quantized with 8 bits and 4 bits, respectively.}  }
    \label{plot2}
\end{figure}
Fig.~\ref{plot2} provides the average rate per user plotted with different number of users. When the BS serves one user at a time (SU-MIMO), i.e., if the resources are orthogonally allocated in Time Division Multiple Access (TDMA) mode , the user gets the highest achievable rate since this scenario is interference-free. When the number of served users, the average rate per user decreases as each user becomes vulnerable to the multiuser interference caused by the other served users. One can observe that TDMA SU-MIMO scenario is better, however, the BS can only serve one user and therefore the resources are not efficiently allocated. Non Orthogonal Multiple Access (NOMA) are introduced in a way to efficiently manage the resources allocated to each users while interference cancellation techniques such as digital precoding are introduced to minimize the multiuser interference.
\begin{figure}[H]
\centering
\setlength\fheight{5.5cm}
\setlength\fwidth{7.5cm}
%
%
\definecolor{mycolor1}{rgb}{0.00000,0.44700,0.74100}%
\definecolor{mycolor2}{rgb}{0.85000,0.32500,0.09800}%
\definecolor{mycolor3}{rgb}{0.92900,0.69400,0.12500}%
\begin{tikzpicture}

\begin{axis}[%
width=0.951\fwidth,
height=\fheight,
at={(0\fwidth,0\fheight)},
scale only axis,
xmin=-20,
xmax=10,
xlabel style={font=\color{white!15!black}},
xlabel={\textsf{SNR (dB)}},
ymin=0,
ymax=8,
ylabel style={font=\color{white!15!black}},
ylabel={\textsf{Average Rate per User (bits/s/Hz)}},
axis background/.style={fill=white},
axis x line*=bottom,
axis y line*=left,
legend style={at={(0.63,0.2)}, anchor=north west, legend cell align=left, align=left, draw=none}
]

\node[right, align=left, rotate=43]
at (axis cs:0,4.2) {\scriptsize\sffamily{SU-MIMO}};

\node[right, align=left, rotate=17]
at (axis cs:-1,2.5) {\scriptsize\sffamily{MU-MIMO (2 Users)}};

\addplot [color=black, dash pattern={on 10pt off 1pt on 0pt off 0pt},line width=1.3pt]
  table[row sep=crcr]{%
-21.2339595381851	0.129471419697568\\
-19.2339595381851	0.20965385657493\\
-17.2339595381851	0.343409191941773\\
-15.2339595381851	0.496857375004632\\
-13.2339595381851	0.698211345497995\\
-11.2339595381851	0.95302739988686\\
-9.2339595381851	1.26156953425137\\
-7.2339595381851	1.61238373751182\\
-5.2339595381851	1.98889143584993\\
-3.2339595381851	2.37874304992816\\
-1.2339595381851	2.75640673440052\\
0.766040461814896	3.08566360963453\\
2.7660404618149	3.37041338042324\\
4.7660404618149	3.60294214303868\\
6.7660404618149	3.76636045574185\\
8.7660404618149	3.90989730796651\\
10.7660404618149	3.96955310619157\\
};
\addlegendentry{$\scriptsize{ \textsf{B}_{\textsf{BB}} = 4~\textsf{bits}    }$}

\addplot [color=black ,line width=1.3pt]
  table[row sep=crcr]{%
-21.2339595381851	0.130284164363507\\
-19.2339595381851	0.212423552985668\\
-17.2339595381851	0.348618537241975\\
-15.2339595381851	0.504835404778792\\
-13.2339595381851	0.714057076099393\\
-11.2339595381851	0.98039464457083\\
-9.2339595381851	1.29486403940671\\
-7.2339595381851	1.65768010114707\\
-5.2339595381851	2.05061519892944\\
-3.2339595381851	2.44245901050533\\
-1.2339595381851	2.82405662890632\\
0.766040461814896	3.16453917753576\\
2.7660404618149	3.46138105638215\\
4.7660404618149	3.69236967911898\\
6.7660404618149	3.89875521363061\\
8.7660404618149	4.07461922342854\\
10.7660404618149	4.23998465806836\\
};
\addlegendentry{$\scriptsize{ \textsf{B}_{\textsf{BB}} = 8~\textsf{bits}    }$}

\addplot [color=black, line width=1.3pt]
  table[row sep=crcr]{%
-21.2339595381851	0.143118736255884\\
-19.2339595381851	0.231387295814562\\
-17.2339595381851	0.383670452201382\\
-15.2339595381851	0.558690232229786\\
-13.2339595381851	0.793284075935042\\
-11.2339595381851	1.09209760646045\\
-9.2339595381851	1.46110232196697\\
-7.2339595381851	1.89719347415485\\
-5.2339595381851	2.38971379341992\\
-3.2339595381851	2.92998420819374\\
-1.2339595381851	3.51188176360326\\
0.766040461814896	4.12155086500618\\
2.7660404618149	4.74724903270545\\
4.7660404618149	5.38914647748775\\
6.7660404618149	6.03841348389946\\
8.7660404618149	6.68859541237598\\
10.7660404618149	7.33854868186856\\
};

\addplot [color=black, line width=0.5pt]
  table[row sep=crcr]{%
2.64142135623731	3.20857864376269\\
2.63492295502918	3.20236600592616\\
2.62815192145801	3.19645168504142\\
2.62112193742753	3.19084763189383\\
2.6138472080903	3.18556517032562\\
2.606342433144	3.18061497435424\\
2.59862277712847	3.17600704660398\\
2.59070383878346	3.17175069809409\\
2.582601619529	3.16785452942446\\
2.57433249113207	3.16432641339679\\
2.56591316262497	3.16117347910642\\
2.55736064654222	3.1584020975369\\
2.54869222454405	3.15601786868644\\
2.53992541249627	3.15402561025225\\
2.53107792507675	3.15242934789569\\
2.5221676399802	3.15123230710775\\
2.51321256179363	3.15043690669152\\
2.50423078561529	3.15004475387458\\
2.49524046049078	3.15005664106137\\
2.48625975274014	3.15047254423203\\
2.47730680925007	3.15129162299091\\
2.46839972080533	3.15251222226472\\
2.45955648553359	3.15413187564689\\
2.45079497253748	3.15614731038128\\
2.44213288578731	3.1585544539753\\
2.43358772834755	3.16134844242901\\
2.42517676700921	3.16452363006354\\
2.41691699739962	3.1680736009291\\
2.40882510964026	3.1719911817693\\
2.40091745462175	3.1762684565159\\
2.39321001096443	3.18089678228436\\
2.38571835273113	3.18586680683813\\
2.37845761795737	3.19116848748626\\
2.37144247806269	3.1967911113762\\
2.36468710820481	3.20272331714075\\
2.35820515863655	3.2089531178554\\
2.35200972712342	3.21546792525975\\
2.34611333247762	3.22225457519404\\
2.34052788926182	3.22929935419929\\
2.33526468371403	3.23658802722745\\
2.33033435094202	3.24410586640548\\
2.32574685343343	3.25183768079529\\
2.3215114609251	3.25976784708926\\
2.31763673167213	3.26788034117961\\
2.31413049515464	3.27615877053746\\
2.31099983625707	3.28458640733652\\
2.30825108095215	3.2931462222541\\
2.30588978351827	3.30182091888156\\
2.30392071531625	3.31059296867425\\
2.30234785514803	3.31944464637065\\
2.30117438121693	3.32835806580893\\
2.30040266470562	3.33731521606869\\
2.30003426498475	3.34629799786473\\
2.300069926462	3.35528826011945\\
2.30050957707795	3.36426783663985\\
2.30135232845161	3.37321858282505\\
2.30259647767559	3.38212241233027\\
2.30423951075704	3.39096133361304\\
2.3062781076976	3.39971748628779\\
2.3087081492019	3.40837317721556\\
2.3115247250013	3.41691091625561\\
2.31472214377576	3.42531345160694\\
2.31829394465408	3.43356380466819\\
2.32223291026902	3.44164530434548\\
2.32653108134113	3.44954162073887\\
2.33117977276167	3.45723679813944\\
2.3361695911422	3.46471528727021\\
2.34149045379533	3.47196197670582\\
2.34713160910836	3.47896222340747\\
2.35308165826849	3.48570188231143\\
2.35932857829595	3.49216733491123\\
2.36585974633816	3.49834551677599\\
2.37266196517618	3.50422394394908\\
2.37972148989171	3.50979073817381\\
2.38702405564075	3.51503465089531\\
2.3945549064779	3.5199450859899\\
2.40229882517301	3.52451212117623\\
2.41024016395991	3.52872652806468\\
2.41836287615506	3.5325797908048\\
2.42665054858238	3.53606412329284\\
2.43508643473851	3.53917248490476\\
2.44365348863171	3.5418985947229\\
2.45233439922588	3.54423694422751\\
2.46111162542011	3.54618280842758\\
2.46996743149315	3.54773225540837\\
2.47888392294118	3.54888215427646\\
2.48784308263633	3.5496301814862\\
2.49682680723304	3.54997482553477\\
2.50581694374862	3.54991539001644\\
2.51479532624415	3.5494519950297\\
2.52374381253146	3.54858557693466\\
2.53264432083225	3.54731788646091\\
2.54147886631503	3.54565148516998\\
2.55022959743622	3.54358974027927\\
2.55887883201191	3.54113681785808\\
2.56740909294741	3.53829767440945\\
2.57580314355238	3.53507804685477\\
2.58404402237018	3.53148444094148\\
2.59211507745106	3.52752411809719\\
2.6	3.52320508075689\\
};

\end{axis}
\end{tikzpicture}%
    \caption{\textcolor{black}{Spectral efficiency performance results: Comparison are made between the SU-MIMO and MIMO-MIMO scenarios. We further illustrate the impacts of the digital quantization for variant II (iterative precoder). Fully-connected structure and mmWave configuration are considered for this simulation.}  }
    \label{plot3}
\end{figure}
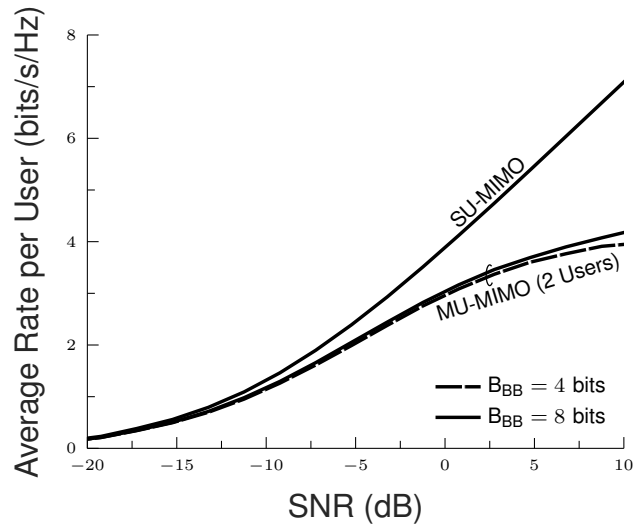
 Fig.~\ref{plot3} illustrates a comparison between SU-MIMO and MIMO-MIMO implemented with 2 users for variant II (iterative precoder). As concluded in Fig.~\ref{plot2}, SU-MIMO outperforms MU-MIMO because of the multiuser interference, however, MU-MIMO is better in terms of resources allocation per user. We also observe that when the digital codebook is oversampled (8 bits), the performance gets much better compared to the regular digital codebook (4 bits). With oversampling, the codebooks contains rich number of beams which improve the chances of selecting the best codewords to improve the spectral efficiency. 
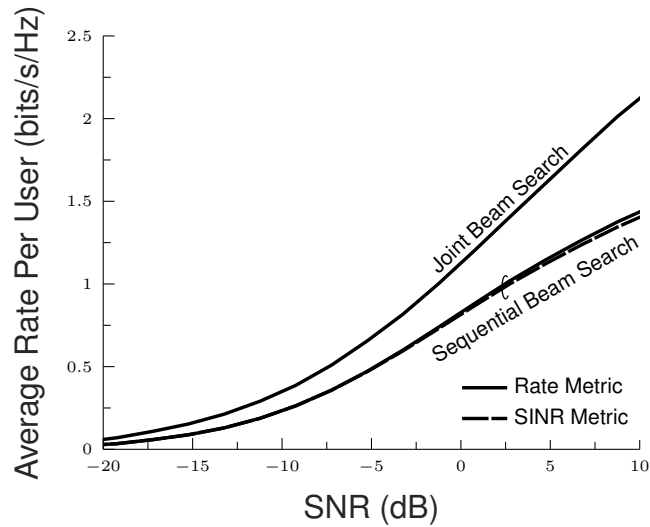
\begin{figure}[H]
\centering
\setlength\fheight{5.5cm}
\setlength\fwidth{7.5cm}
%
%
\definecolor{mycolor1}{rgb}{0.00000,0.44700,0.74100}%
\definecolor{mycolor2}{rgb}{0.85000,0.32500,0.09800}%
\definecolor{mycolor3}{rgb}{0.92900,0.69400,0.12500}%
\begin{tikzpicture}

\begin{axis}[%
width=0.951\fwidth,
height=\fheight,
at={(0\fwidth,0\fheight)},
scale only axis,
xmin=-20,
xmax=10,
xlabel style={font=\color{white!15!black}},
xlabel={\textsf{SNR (dB)}},
ymin=0,
ymax=2.5,
ylabel style={font=\color{white!15!black}},
ylabel={\textsf{Average Rate Per User (bits/s/Hz)}},
axis background/.style={fill=white},
axis x line*=bottom,
axis y line*=left,
legend style={at={(0.65,0.2)}, anchor=north west, legend cell align=left, align=left, draw=none}]

\node[right, align=left, rotate=42.7]
at (axis cs:-2,1.05) {\scriptsize\sffamily{Joint Beam Search}};

\node[right, align=left, rotate=30]
at (axis cs:-2,0.52) {\scriptsize\sffamily{Sequential Beam Search}};

\addplot [color=black, line width=1.3pt]
  table[row sep=crcr]{%
-21.2339595381851	0.0205632647347906\\
-19.2339595381851	0.0341471689982687\\
-17.2339595381851	0.0584902764688735\\
-15.2339595381851	0.0879778853371111\\
-13.2339595381851	0.129695595602693\\
-11.2339595381851	0.187448614371497\\
-9.2339595381851	0.263135972359829\\
-7.2339595381851	0.358041549964817\\
-5.2339595381851	0.472045095563663\\
-3.2339595381851	0.600543699006455\\
-1.2339595381851	0.739743677737465\\
0.766040461814896	0.881696116590055\\
2.7660404618149	1.02092665220472\\
4.7660404618149	1.14734887609477\\
6.7660404618149	1.26536445601709\\
8.7660404618149	1.37595824855567\\
10.7660404618149	1.47315812177755\\
};
\addlegendentry{\scriptsize{\textsf{Rate Metric}}}

\addplot [color=black,dash pattern={on 10pt off 1pt on 0pt off 0pt}, line width=1.3pt]
  table[row sep=crcr]{%
-21.2339595381851	0.0205632646488328\\
-19.2339595381851	0.0341471309397087\\
-17.2339595381851	0.0584892488826138\\
-15.2339595381851	0.0879710333872069\\
-13.2339595381851	0.129658929460885\\
-11.2339595381851	0.187332682591742\\
-9.2339595381851	0.262733150617674\\
-7.2339595381851	0.3569594507253\\
-5.2339595381851	0.469714578156892\\
-3.2339595381851	0.595547386033713\\
-1.2339595381851	0.731059246763519\\
0.766040461814896	0.868037032465312\\
2.7660404618149	1.00211682196704\\
4.7660404618149	1.12376148485179\\
6.7660404618149	1.23781062295286\\
8.7660404618149	1.34508369411328\\
10.7660404618149	1.43984783236813\\
};
\addlegendentry{\scriptsize{\textsf{SINR Metric}}}

\addplot [color=black, line width=1.3pt, forget plot]
  table[row sep=crcr]{%
-21.2339595381851	0.0437859003709619\\
-19.2339595381851	0.0699274089355573\\
-17.2339595381851	0.10851026237436\\
-15.2339595381851	0.153239666707829\\
-13.2339595381851	0.212742560651434\\
-11.2339595381851	0.290655788339489\\
-9.2339595381851	0.386767108778771\\
-7.2339595381851	0.509180558353936\\
-5.2339595381851	0.655813924293522\\
-3.2339595381851	0.817468745546691\\
-1.2339595381851	1.00298863923154\\
0.766040461814896	1.20257438515081\\
2.7660404618149	1.4084104336842\\
4.7660404618149	1.61156180398803\\
6.7660404618149	1.81474872751652\\
8.7660404618149	2.01360270932133\\
10.7660404618149	2.19004464200282\\
};
\addplot [color=black, line width=.5pt]
  table[row sep=crcr]{%
2.64142135623731	0.930502525316942\\
2.63492295502918	0.928328102074157\\
2.62815192145801	0.926258089764496\\
2.62112193742753	0.924296671162842\\
2.6138472080903	0.922447809613966\\
2.606342433144	0.920715241023984\\
2.59862277712847	0.919102466311394\\
2.59070383878346	0.917612744332932\\
2.582601619529	0.916249085298559\\
2.57433249113207	0.915014244688875\\
2.56591316262497	0.913910717687247\\
2.55736064654222	0.912940734137916\\
2.54869222454405	0.912106254040254\\
2.53992541249627	0.911408963588288\\
2.53107792507675	0.91085027176349\\
2.5221676399802	0.910431307487712\\
2.51321256179363	0.910152917342032\\
2.50423078561529	0.910015663856103\\
2.49524046049078	0.910019824371481\\
2.48625975274014	0.910165390481212\\
2.47730680925007	0.910452068046818\\
2.46839972080533	0.910879277792652\\
2.45955648553359	0.911446156476411\\
2.45079497253748	0.912151558633447\\
2.44213288578731	0.912994058891355\\
2.43358772834755	0.913971954850153\\
2.42517676700921	0.91508327052224\\
2.41691699739962	0.916325760325184\\
2.40882510964026	0.917696913619256\\
2.40091745462175	0.919193959780566\\
2.39321001096443	0.920813873799526\\
2.38571835273113	0.922553382393344\\
2.37845761795737	0.92440897062019\\
2.37144247806269	0.926376888981671\\
2.36468710820481	0.928453160999263\\
2.35820515863655	0.930633591249388\\
2.35200972712342	0.932913773840912\\
2.34611333247762	0.935289101317913\\
2.34052788926182	0.93775477396975\\
2.33526468371403	0.940305809529606\\
2.33033435094202	0.942937053241919\\
2.32574685343343	0.94564318827835\\
2.3215114609251	0.948418746481241\\
2.31763673167213	0.951258119412862\\
2.31413049515464	0.954155569688112\\
2.31099983625707	0.957105242567781\\
2.30825108095215	0.960101177788936\\
2.30588978351827	0.963137321608547\\
2.30392071531625	0.966207539035989\\
2.30234785514803	0.969305626229728\\
2.30117438121693	0.972425323033126\\
2.30040266470562	0.97556032562404\\
2.30003426498475	0.978704299252656\\
2.300069926462	0.981850891041809\\
2.30050957707795	0.984993742823946\\
2.30135232845161	0.988126503988766\\
2.30259647767559	0.991242844315596\\
2.30423951075704	0.994336466764563\\
2.3062781076976	0.997401120200728\\
2.3087081492019	1.00043061202545\\
2.3115247250013	1.00341882068946\\
2.31472214377576	1.00635970806243\\
2.31829394465408	1.00924733163387\\
2.32223291026902	1.01207585652092\\
2.32653108134113	1.01483956725861\\
2.33117977276167	1.0175328793488\\
2.3361695911422	1.02015035054457\\
2.34149045379533	1.02268669184704\\
2.34713160910836	1.02513677819262\\
2.35308165826849	1.027495658809\\
2.35932857829595	1.02975856721893\\
2.36585974633816	1.0319209308716\\
2.37266196517618	1.03397838038218\\
2.37972148989171	1.03592675836083\\
2.38702405564075	1.03776212781336\\
2.3945549064779	1.03948078009647\\
2.40229882517301	1.04107924241168\\
2.41024016395991	1.04255428482264\\
2.41836287615506	1.04390292678168\\
2.42665054858238	1.04512244315249\\
2.43508643473851	1.04621036971667\\
2.44365348863171	1.04716450815301\\
2.45233439922588	1.04798293047963\\
2.46111162542011	1.04866398294965\\
2.46996743149315	1.04920628939293\\
2.47888392294118	1.04960875399676\\
2.48784308263633	1.04987056352017\\
2.49682680723304	1.04999118893717\\
2.50581694374862	1.04997038650575\\
2.51479532624415	1.0498081982604\\
2.52374381253146	1.04950495192713\\
2.53264432083225	1.04906126026132\\
2.54147886631503	1.04847801980949\\
2.55022959743622	1.04775640909774\\
2.55887883201191	1.04689788625033\\
2.56740909294741	1.04590418604331\\
2.57580314355238	1.04477731639917\\
2.58404402237018	1.04351955432952\\
2.59211507745106	1.04213344133402\\
2.6	1.04062177826491\\
};

\end{axis}
\end{tikzpicture}%
    \caption{\textcolor{black}{Spectral efficiency performance results: Comparison are made between the sequential and joint beam search which serves as a benchmarking tool. We further illustrate the impacts of the digital quantization for variant II (iterative precoder). Note that partially connected structure and mmWave configuration are considered while 3 users are simultaneously served at a time. }}
    \label{plot4}
\end{figure}
Fig.~\ref{plot4} illustrates a comparison between the sequential and joint beam search which serves as a benchmarking tool. We observe that the joint (exhaustive) beam search approach outperforms the sequential (two-stage) beam search, however, the two-stage method is practical as it requires low-complexity cost. We further note that the design based on the rate metric is roughly better than the performance derived for the SINR metric.  
\begin{figure}[H]
\centering
\setlength\fheight{5.5cm}
\setlength\fwidth{7.5cm}
%
%
\definecolor{mycolor1}{rgb}{0.00000,0.44700,0.74100}%
\definecolor{mycolor2}{rgb}{0.85000,0.33000,0.10000}%
\begin{tikzpicture}

\begin{axis}[%
width=0.951\fwidth,
height=\fheight,
at={(0\fwidth,0\fheight)},
scale only axis,
xmin=-10,
xmax=10,
axis x line*=bottom,
axis y line*=left,
xlabel style={font=\color{white!15!black}},
xlabel={\sffamily{SNR (dB)}},
ymode=log,
ymin=1e-09,
ymax=1,
yminorticks=true,
ylabel style={font=\color{white!15!black}},
ylabel={\sffamily{Bit Error Probability}},
axis background/.style={fill=white},
legend style={at={(0.97,0.97)}, anchor=north east, legend cell align=left, align=left, draw=white!15!black,draw=none,fill=none}
]
\node[right, align=left, rotate=-47]
at (axis cs:-8,1e-3) {\scriptsize\textsf{Static ($v_m = 0$ km/h)}};

\node[right, align=left, rotate=-47]
at (axis cs:-3.8,1e-3) {\scriptsize\textsf{Moderate Mobility ($v_m = 50$ km/h)}};

\node[right, align=left, rotate=-47]
at (axis cs:1,1e-3) {\scriptsize\textsf{High Mobility ($v_m = 90$ km/h)}};

\addplot [color=black, line width=1.3pt]
  table[row sep=crcr]{%
-11.5204205106835	0.011134550113225\\
-10.5204205106835	0.00501102265463745\\
-9.52042051068354	0.00193428037855052\\
-8.52042051068354	0.000640986210067648\\
-7.52042051068354	0.000183106774865973\\
-6.52042051068354	4.5417984297775e-05\\
-5.52042051068354	9.87894487066316e-06\\
-4.52042051068354	1.90670585786826e-06\\
-3.52042051068354	3.30797374744084e-07\\
-2.52042051068354	5.22759521252148e-08\\
-1.52042051068354	7.62280233553235e-09\\
-0.520420510683537	1.0381388995365e-09\\
0.479579489316461	1.33502821644045e-10\\
1.47957948931646	1.63692926580519e-11\\
2.47957948931646	1.92984825155453e-12\\
3.47957948931646	2.20301397150991e-13\\
4.47957948931646	2.44983461761851e-14\\
5.47957948931646	2.69925048285598e-15\\
6.47957948931646	2.44424630270574e-16\\
7.47957948931646	6.38383764902322e-17\\
8.47957948931646	8.56286129891127e-18\\
9.47957948931646	4.67126538924097e-17\\
10.4795794893165	8.56286129891127e-18\\
11.4795794893165	1.90748962967492e-17\\
12.4795794893165	2.95869312945872e-17\\
13.4795794893165	-3.81497925934984e-17\\
14.4795794893165	1.05120349978379e-17\\
15.4795794893165	1.71257225978225e-17\\
16.4795794893165	3.89834739785334e-18\\
17.4795794893165	0\\
18.4795794893165	1.90748962967492e-17\\
};

\addplot [color=black,  line width=1.3pt]
  table[row sep=crcr]{%
-11.5204205106835	0.0463046804863404\\
-10.5204205106835	0.035121688621787\\
-9.52042051068354	0.0243529098714744\\
-8.52042051068354	0.0152263403427127\\
-7.52042051068354	0.00847416746124228\\
-6.52042051068354	0.00415103357781563\\
-5.52042051068354	0.00177403886957403\\
-4.52042051068354	0.000657839653127797\\
-3.52042051068354	0.000211277323363118\\
-2.52042051068354	5.88934543021557e-05\\
-1.52042051068354	1.4329648882945e-05\\
-0.520420510683537	3.07002709823008e-06\\
0.479579489316461	5.8550314029078e-07\\
1.47957948931646	1.00626254554647e-07\\
2.47957948931646	1.57838103328117e-08\\
3.47957948931646	2.28802331123803e-09\\
4.47957948931646	3.10148820115047e-10\\
5.47957948931646	3.97369604785362e-11\\
6.47957948931646	4.85805957684542e-12\\
7.47957948931646	5.71387273267652e-13\\
8.47957948931646	6.51337875534202e-14\\
9.47957948931646	7.23756697127895e-15\\
10.4795794893165	8.22034684695482e-16\\
11.4795794893165	6.85028903912902e-17\\
12.4795794893165	-2.10240699956759e-17\\
13.4795794893165	-1.24612086967646e-17\\
14.4795794893165	1.05120349978379e-17\\
15.4795794893165	1.05120349978379e-17\\
16.4795794893165	1.05120349978379e-17\\
17.4795794893165	-6.6136875999846e-18\\
18.4795794893165	4.00989662924251e-17\\
};

\addplot [color=black, line width=1.3pt]
  table[row sep=crcr]{%
-11.5204205106835	0.0707469802111243\\
-10.5204205106835	0.0667301362562948\\
-9.52042051068354	0.0612180407797725\\
-8.52042051068354	0.0541373642237648\\
-7.52042051068354	0.0456672993194797\\
-6.52042051068354	0.0363081290601155\\
-5.52042051068354	0.0268524400576826\\
-4.52042051068354	0.0182189737928771\\
-3.52042051068354	0.0111832001094422\\
-2.52042051068354	0.00612857922674826\\
-1.52042051068354	0.00296381974647623\\
-0.520420510683537	0.00125336056784779\\
0.479579489316461	0.000460772306977135\\
1.47957948931646	0.000146951236755322\\
2.47957948931646	4.07307387848857e-05\\
3.47957948931646	9.86513806326895e-06\\
4.47957948931646	2.10577492683602e-06\\
5.47957948931646	4.0042161660818e-07\\
6.47957948931646	6.8655270919927e-08\\
7.47957948931646	1.07486691821366e-08\\
8.47957948931646	1.55578286138039e-09\\
9.47957948931646	2.10638073414896e-10\\
10.4795794893165	2.69616291505568e-11\\
11.4795794893165	3.29364932311103e-12\\
12.4795794893165	3.87152647907675e-13\\
13.4795794893165	4.41415499958876e-14\\
14.4795794893165	4.92559442057291e-15\\
15.4795794893165	5.30897400532499e-16\\
16.4795794893165	8.56286129891127e-17\\
17.4795794893165	-2.10240699956759e-17\\
18.4795794893165	-2.10240699956759e-17\\
};

\end{axis}
\end{tikzpicture}%
    \caption{\textcolor{black}{Bit error probability performance results: We investigate the effects of the vehicles mobility on the error performance. In this scenario, we consider MIMO-OFDM multicarrier transmission and QPSK symbols mapped from coded bits. As in 5G NR standards, we consider LDPC channel coding with 1/2 coding rate. Note that partially-connected structure and sub-6 GHz configuration are assumed. }}
    \label{plot5}
\end{figure}
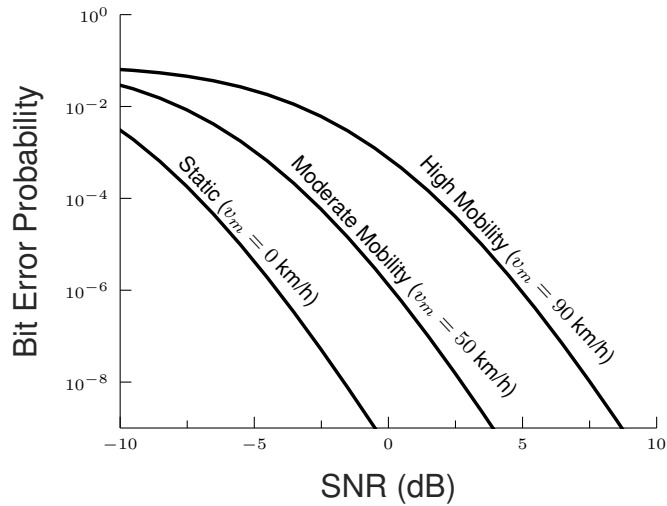
Fig.~\ref{plot5} provides the error performance against the average SNR for different values of vehicles velocity. We observe that the error rate increases with the velocity and vice versa. In fact, the mobility introduces the time selectivity to the channel and hence the Doppler shift which in turn creates the CFO. If the CFO is not perfectly estimated and canceled, it rotates the phases of the constellation points resulting in worse error performance. Various algorithms have been developed to estimate and compensate for the CFO before decoding which is out of the scope of this work. For example, Moose and Schmidl-Cox algorithms have been proposed to estimate the CFO using periodic training for OFDM systems \cite{cfo}.
\begin{figure}[H]
\centering
\setlength\fheight{5.5cm}
\setlength\fwidth{7.5cm}
\input{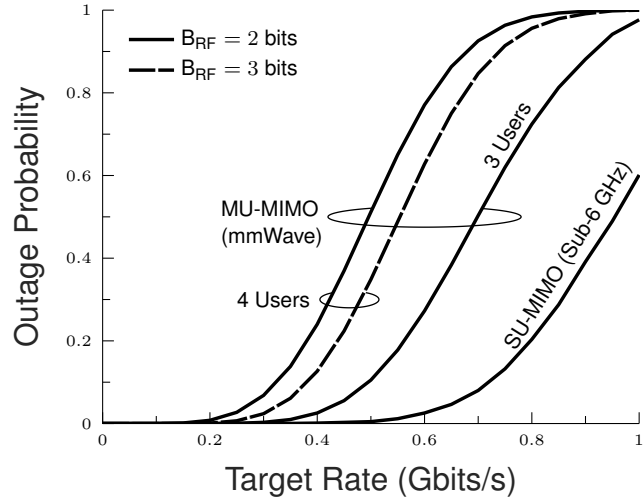}
    \caption{\textcolor{black}{Outage probability performance results: We consider MU-MIMO for 3 and 4 users implemented in mmWave configuration againts SU-MIMO implemented in sub-6 GHz setting. We further consider different number of bits to quantize the analog beam codebooks for variant II. The digital codebook is quantized using 2 bits while the SNR is fixed at 0 dB.}}
    \label{plot6}
\end{figure}  
Fig.~\ref{plot6} illustrates the variations of the outage probability against the target rate for MU-MIMO and SU-MIMO scenarios. Table \ref{services} shows the difference use cases services requirements for 5G V2X systems. According to Fig.~\ref{plot6}, all the users can support the following services (vehicle platooning, remote driving, cooperative collision avoidance and info sharing) with null probability of outage, i.e., all the users are in coverage for these services. However, each user considered for MU-MIMO (3 and 4 users) can support the video data sharing service with 50$\%$ and 95$\%$ (2 bits quantization) of outage probability. We also observe that the user considered for SU-MIMO scenario has 40$\%$ coverage to support the collective perception of environment service. We also note that the performance gets much better by oversampling the analog beam codebooks (3 bits) compared to the regular analog codebook (2 bits).
\begin{figure}[H]
\centering
\setlength\fheight{5.5cm}
\setlength\fwidth{7.5cm}
\input{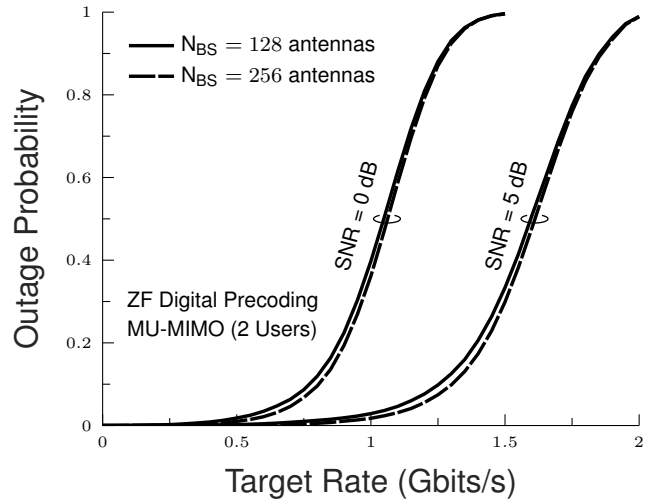}
    \caption{\textcolor{black}{Outage probability performance results: We consider the massive MU-MIMO scenario with 128 and 256 antennas at the BS implemented with full-connected structure and mmWave setting. Besides, we consider the ZF digital precoding to serve 2 users while the SNR is fixed at 0 and 5 dB.}}
    \label{plot7}
\end{figure}
Fig.~\ref{plot7} illustrates the variation of the outage probability against the target rate evaluated for 0 and 5 dB and for massive number of antennas at the BS. We observe that by increasing the number of antennas, the outage probability is improved, however, this enhancement is negligible. In fact, the first stage analog beamforming is implemented in SU-MIMO scenario, i.e., the analog beamformers are designed to increased the received power without accounting for the multiuser interference. Therefore, the best analog beamformers selected in the first stage are not necessarily robust to minimize the multiuser interference. In addition, the outage performance is enhanced with the average SNR as expected.
\begin{figure}[H]
\centering
\setlength\fheight{5.5cm}
\setlength\fwidth{7.5cm}
%
%
\begin{tikzpicture}

\begin{axis}[%
width=0.951\fwidth,
height=\fheight,
at={(0\fwidth,0\fheight)},
scale only axis,
xmin=0,
xmax=14,
xlabel style={font=\color{white!15!black}},
xlabel={\textsf{Number of Users}},
ymin=0,
ymax=6,
ylabel style={font=\color{white!15!black}},
ylabel={\textsf{Sum Energy Efficiency (Gbits/s/Joule)}},
axis background/.style={fill=white},
axis x line*=bottom,
axis y line*=left,
legend style={legend cell align=left, align=left, draw=none,fill=none}
]

\node[right, align=left, rotate=0]
at (axis cs:1,1) {\scriptsize{\textsf{MRT Digital Precoding}}};

\addplot [color=black, line width=1.3pt]
  table[row sep=crcr]{%
1	4.40577356349385\\
2	5.52086572332713\\
3	5.50687584487464\\
4	5.03681193848618\\
5	4.18983928053848\\
6	3.05568706690332\\
7	2.06331440521361\\
8	1.29510753033586\\
9	0.771476863660851\\
10	0.441345517699016\\
11	0.235598651002849\\
12	0.124037127699909\\
13	0.061599562807192\\
14	0.0229163013698764\\
};
\addlegendentry{\scriptsize\textsf{Mode 1}}

\addplot [color=black, dash pattern={on 10pt off 1pt on 0pt off 0pt}, line width=1.0pt]
  table[row sep=crcr]{%
1	2.61255720132411\\
2	3.60011469252724\\
3	3.88559394592794\\
4	3.74707512412952\\
5	3.54182475216733\\
6	3.27891040220639\\
7	3.04939910297844\\
8	2.72675095373126\\
9	2.41588002083983\\
10	2.1347069483329\\
11	1.83719141707841\\
12	1.53922889975128\\
13	1.23689975114746\\
14	1.02395862566205\\
};
\addlegendentry{\scriptsize\textsf{Mode 2}}

\addplot [color=black, dash pattern={on 10pt off 1pt on 1pt off 1pt}, line width=1.0pt]
  table[row sep=crcr]{%
1	2.5330332029839\\
2	3.38395373423599\\
3	3.5765545915764\\
4	3.55033434845953\\
5	3.35055012359813\\
6	3.06940876503319\\
7	2.7794691453853\\
8	2.50164450304218\\
9	2.22320762999804\\
10	1.90673765678991\\
11	1.67079809791451\\
12	1.39951045049681\\
13	1.17197758529628\\
14	0.9526463643527\\
};
\addlegendentry{\scriptsize\textsf{Mode 3}}

\addplot [color=black, dotted, line width=1.3pt]
  table[row sep=crcr]{%
1	1.58235599179846\\
2	2.29965538636777\\
3	2.61145416607309\\
4	2.7104871782125\\
5	2.63279188818791\\
6	2.55166629706988\\
7	2.43410095942176\\
8	2.2091627823809\\
9	2.08415038372318\\
10	1.9326146349194\\
11	1.79548901026595\\
12	1.69988116960104\\
13	1.58983054814464\\
14	1.45347361163997\\
};
\addlegendentry{\scriptsize\textsf{Mode 4}}

\end{axis}
\end{tikzpicture}%
    \caption{\textcolor{black}{Energy efficiency performance: Results are evaluated for four modes and for different number of users. We consider the MRT digital precoding and 0 dB of SNR. }}
    \label{plot8}
\end{figure}
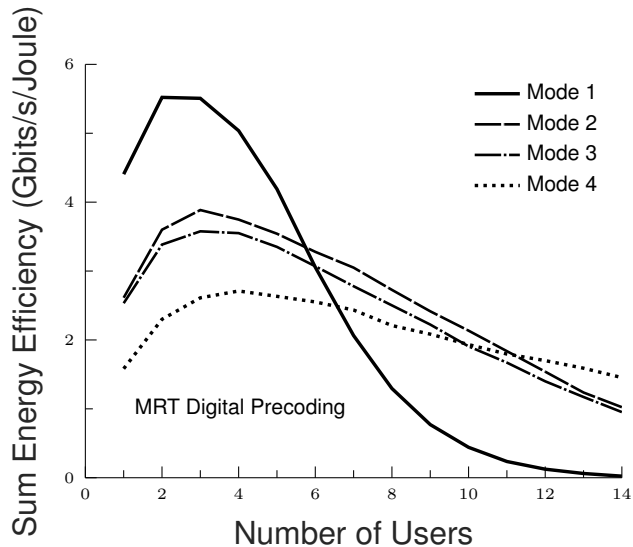
In this simulation, we will evaluate the energy efficiency for different variants wherein the relative parameters of each one is provided by Table \ref{setting}.
\begin{table}[H]
\renewcommand{\arraystretch}{.7}
\caption{Parameters of Each Mode \cite{ni,heath}.}
\label{setting}
\centering
\textcolor{black}{
\begin{tabular}{rllll}
\bfseries Parameter & \bfseries Mode 1 & \bfseries Mode 2 & \bfseries Mode 3 & \bfseries Mode 4\\
\hline
 Frequency & 28 GHz & 39 GHz & 39 GHz & 73 GHz\\
 Bandwidth & 850 MHz & 1.4 GHz & 1.6 GHz & 2 GHz\\
 ADC Power & 250 mW & 400 mW & 450 mW & 550 mW\\
BS Antennas & 32 & 64 & 64 & 128\\
\end{tabular}}
\end{table}
Fig. \ref{plot8} illustrates the energy efficiency as a function of the number of users for different modes. We observe that for low number of users, mode 1 seems to be well advocated for power-efficient systems as it provides better energy efficiency compared to the other modes. This is explained by the fact that for a few number of users, the rate at which the power consumption increases is lower than the rate of spectral efficiency. However, as the number of users increases, the spectral efficiency improves at lower rate compared to the power consumption as a function of the number of users wherein mode 4 becomes the best one since it provides hefty beamforming gain achieved by the massive number of antennas at the BS. Furthermore, the effect of the ADC power can be easily observed by comparing modes 2 and 3 (having the same number of antennas at the BS). We notice that mode 2 is more power-efficient than mode 3 since the latter requires more ADC power. We also note that mode 2 remains more power-efficient than mode 3 across all the number of users.}

\section{Conclusion}
\textcolor{black}{In this work, we proposed a low-complexity hybrid precoding algorithm for downlink V2X multiuser mmWave systems. We developed closed-form digital precoder such as MRT, ZF and MMSE in variant I while iterative precoder was designed in variant II which are required to manage the multiuser interference. We observed that when the feedback is limited, the joint analog/digital codebooks quantization incurred rate loss which grows with the number of users. The results also indicated that the mobility has a destructive effects on the error performance due to the CFO incurred by the Doppler shift. Moreover, we investigated the ability of MU-MIMO system to support different V2X services for different number users as well as for mmWave and sub-6 GHz configurations. Besides, we studied the effects of the number of users as well as the ADC power on the energy efficiency for different mmWave modes. Estimating the sidelink V2V rate is important to quantify the reliability of exchanging the data and support V2V services. This requirement is accomplished by estimating the V2X uplink spectral efficiency which requires robust power control unlike the downlink scenario investigated in this work. The analysis of the uplink MU-MIMO V2X framework is deferred to the next future direction. In addition, it is of interest to extend this work to support multicell cellular systems and develop efficient MU-MIMO hybrid precoding and channel estimation algorithms taking into consideration the out-of-cell interference. Finally, it is important to consider the blockage effects while designing the beamformers which deserves careful attention and we deferred this analysis to the next future direction.}

\ifCLASSOPTIONcaptionsoff
  \newpage
\fi
\bibliographystyle{IEEEtran}
\bibliography{main}

\begin{thebibliography}{10}
\providecommand{\url}[1]{#1}
\csname url@samestyle\endcsname
\providecommand{\newblock}{\relax}
\providecommand{\bibinfo}[2]{#2}
\providecommand{\BIBentrySTDinterwordspacing}{\spaceskip=0pt\relax}
\providecommand{\BIBentryALTinterwordstretchfactor}{4}
\providecommand{\BIBentryALTinterwordspacing}{\spaceskip=\fontdimen2\font plus
\BIBentryALTinterwordstretchfactor\fontdimen3\font minus
  \fontdimen4\font\relax}
\providecommand{\BIBforeignlanguage}[2]{{%
\expandafter\ifx\csname l@#1\endcsname\relax
\typeout{** WARNING: IEEEtran.bst: No hyphenation pattern has been}%
\typeout{** loaded for the language `#1'. Using the pattern for}%
\typeout{** the default language instead.}%
\else
\language=\csname l@#1\endcsname
\fi
#2}}
\providecommand{\BIBdecl}{\relax}
\BIBdecl

\bibitem{v1}
E.~Ohn-Bar and M.~M. Trivedi, ``{Looking at Humans in the Age of Self-Driving
  and Highly Automated Vehicles},'' \emph{IEEE Transactions on Intelligent
  Vehicles}, vol.~1, no.~1, pp. 90--104, March 2016.

\bibitem{v2}
J.~Pimentel and J.~Bastiaan, ``{Characterizing the Safety of Self-Driving
  Vehicles: A Fault Containment Protocol for Functionality Involving Vehicle
  Detection},'' in \emph{2018 IEEE International Conference on Vehicular
  Electronics and Safety (ICVES)}, Sep. 2018, pp. 1--7.

\bibitem{v3}
F.~Jimenez, J.~E. Naranjo, J.~J. Anaya, F.~Garcia, A.~Ponz, and J.~M. Armingol,
  ``{Advanced Driver Assistance System for Road Environments to Improve Safety
  and Efficiency},'' \emph{Transportation Research Procedia}, vol.~14, pp. 2245
  -- 2254, 2016, transport Research Arena TRA2016.

\bibitem{va}
J.~{Choi}, V.~{Va}, N.~{Gonzalez-Prelcic}, R.~{Daniels}, C.~R. {Bhat}, and
  R.~W. {Heath}, ``{Millimeter-Wave Vehicular Communication to Support Massive
  Automotive Sensing},'' \emph{IEEE Communications Magazine}, vol.~54, no.~12,
  pp. 160--167, December 2016.

\bibitem{lte}
P.~{Belanovic}, D.~{Valerio}, A.~{Paier}, T.~{Zemen}, F.~{Ricciato}, and C.~F.
  {Mecklenbrauker}, ``{On Wireless Links for Vehicle-to-Infrastructure
  Communications},'' \emph{IEEE Transactions on Vehicular Technology}, vol.~59,
  no.~1, pp. 269--282, Jan 2010.

\bibitem{tractable}
E.~{Balti} and B.~K. {Johnson}, ``Tractable approach to mmwaves cellular
  analysis with fso backhauling under feedback delay and hardware
  limitations,'' \emph{IEEE Transactions on Wireless Communications}, vol.~19,
  no.~1, pp. 410--422, 2020.

\bibitem{decade}
K.~{Sato} and M.~{Fujise}, ``{Propagation Measurements for Inter-Vehicle
  Communication in 76-GHz Band},'' in \emph{2006 6th International Conference
  on ITS Telecommunications}, June 2006, pp. 408--411.

\bibitem{surv}
V.~Va, T.~Shimizu, G.~Bansal, and R.~W. Heath, Jr., \emph{{Millimeter Wave
  Vehicular Communications: A Survey}}.\hskip 1em plus 0.5em minus 0.4em\relax
  Hanover, MA, USA: Now Publishers Inc., 2016.

\bibitem{tassi}
A.~{Tassi}, M.~{Egan}, R.~J. {Piechocki}, and A.~{Nix}, ``{Modeling and Design
  of Millimeter-Wave Networks for Highway Vehicular Communication},''
  \emph{IEEE Transactions on Vehicular Technology}, vol.~66, no.~12, pp.
  10\,676--10\,691, Dec 2017.

\bibitem{urban}
Y.~{Wang}, K.~{Venugopal}, A.~F. {Molisch}, and R.~W. {Heath}, ``{MmWave
  Vehicle-to-Infrastructure Communication: Analysis of Urban Microcellular
  Networks},'' \emph{IEEE Transactions on Vehicular Technology}, vol.~67,
  no.~8, pp. 7086--7100, Aug 2018.

\bibitem{inv}
V.~{Va}, J.~{Choi}, T.~{Shimizu}, G.~{Bansal}, and R.~W. {Heath}, ``{Inverse
  Multipath Fingerprinting for Millimeter Wave V2I Beam Alignment},''
  \emph{IEEE Transactions on Vehicular Technology}, vol.~67, no.~5, pp.
  4042--4058, May 2018.

\bibitem{pos}
V.~{Va}, T.~{Shimizu}, G.~{Bansal}, and R.~W. {Heath}, ``{Position-aided
  millimeter wave V2I beam alignment: A learning-to-rank approach},'' in
  \emph{2017 IEEE 28th Annual International Symposium on Personal, Indoor, and
  Mobile Radio Communications (PIMRC)}, Oct 2017, pp. 1--5.

\bibitem{prediction}
Y.~{Wang}, M.~{Narasimha}, and R.~W. {Heath}, ``{MmWave Beam Prediction with
  Situational Awareness: A Machine Learning Approach},'' in \emph{2018 IEEE
  19th International Workshop on Signal Processing Advances in Wireless
  Communications (SPAWC)}, June 2018, pp. 1--5.

\bibitem{spatially}
A.~F. {Molisch}, A.~{Karttunen}, S.~{Hur}, J.~{Park}, and J.~{Zhang},
  ``{Spatially consistent pathloss modeling for millimeter-wave channels in
  urban environments},'' in \emph{2016 10th European Conference on Antennas and
  Propagation (EuCAP)}, April 2016, pp. 1--5.

\bibitem{cross}
Y.~{Wang}, K.~{Venugopal}, A.~F. {Molisch}, and R.~W. {Heath}, ``{Blockage and
  Coverage Analysis with MmWave Cross Street BSs Near Urban Intersections},''
  in \emph{2017 IEEE International Conference on Communications (ICC)}, May
  2017, pp. 1--6.

\bibitem{relay}
E.~{Balti} and M.~{Guizani}, ``Impact of non-linear high-power amplifiers on
  cooperative relaying systems,'' \emph{IEEE Transactions on Communications},
  vol.~65, no.~10, pp. 4163--4175, 2017.

\bibitem{zeroforcing}
E.~{Balti} and N.~{Mensi}, ``Zero-forcing max-power beamforming for hybrid
  mmwave full-duplex mimo systems,'' in \emph{2020 4th International Conference
  on Advanced Systems and Emergent Technologies (IC\_ASET)}, 2020, pp.
  344--349.

\bibitem{aggregate}
E.~{Balti}, M.~{Guizani}, B.~{Hamdaoui}, and B.~{Khalfi}, ``Aggregate hardware
  impairments over mixed rf/fso relaying systems with outdated csi,''
  \emph{IEEE Transactions on Communications}, vol.~66, no.~3, pp. 1110--1123,
  2018.

\bibitem{surface5g}
C.~{Pradhan}, A.~{Li}, L.~{Song}, B.~{Vucetic}, and Y.~{Li}, ``Hybrid precoding
  design for reconfigurable intelligent surface aided mmwave communication
  systems,'' \emph{IEEE Wireless Communications Letters}, vol.~9, no.~7, pp.
  1041--1045, 2020.

\bibitem{surface6g}
H.~{Hashida}, Y.~{Kawamoto}, and N.~{Kato}, ``Intelligent reflecting surface
  placement optimization in air-ground communication networks toward 6g,''
  \emph{IEEE Wireless Communications}, vol.~27, no.~6, pp. 146--151, 2020.

\bibitem{3gpp2}
G.~W.~G. 1, ``Ls on prioritised use cases and requirements for consideration in
  rel-16 nr-v2x,'' R1-1809720, Tech. Rep., Aug 2018.

\bibitem{3gpp3}
G.~T. 22.886, ``Study on enhancement of 3gpp support for 5g v2x services
  (release 16),'' Technical Report V16.2.0, Tech. Rep., Dec 2018.

\bibitem{3gpp4}
5G-PPP, ``5g empowering vertical industries,'' 5GPPP White Paper, Tech. Rep.,
  Feb 2016.

\bibitem{3gpp5}
G.~T. 38.885, ``Nr: Study on vehicle-to-everything,'' Tech. Rep., Nov 2018.

\bibitem{3gpp6}
G.~T. 37.885, ``Methodology of new v2x use cases for lte and nr,'' Release 15,
  Tech. Rep., Dec 2018.

\bibitem{ayach}
O.~E. {Ayach}, S.~{Rajagopal}, S.~{Abu-Surra}, Z.~{Pi}, and R.~W. {Heath},
  ``Spatially sparse precoding in millimeter wave mimo systems,'' \emph{IEEE
  Transactions on Wireless Communications}, vol.~13, no.~3, pp. 1499--1513,
  2014.

\bibitem{multibeam}
C.~{Kim}, T.~{Kim}, and J.~{Seol}, ``Multi-beam transmission diversity with
  hybrid beamforming for mimo-ofdm systems,'' in \emph{2013 IEEE Globecom
  Workshops (GC Wkshps)}, 2013, pp. 61--65.

\bibitem{phase}
{Xinying Zhang}, A.~F. {Molisch}, and {Sun-Yuan Kung},
  ``Variable-phase-shift-based rf-baseband codesign for mimo antenna
  selection,'' \emph{IEEE Transactions on Signal Processing}, vol.~53, no.~11,
  pp. 4091--4103, 2005.

\bibitem{estimation}
V.~{Venkateswaran} and A.~{van der Veen}, ``Analog beamforming in mimo
  communications with phase shift networks and online channel estimation,''
  \emph{IEEE Transactions on Signal Processing}, vol.~58, no.~8, pp.
  4131--4143, 2010.

\bibitem{alkhateeb}
A.~{Alkhateeb}, G.~{Leus}, and R.~W. {Heath}, ``Limited feedback hybrid
  precoding for multi-user millimeter wave systems,'' \emph{IEEE Transactions
  on Wireless Communications}, vol.~14, no.~11, pp. 6481--6494, 2015.

\bibitem{feedback}
D.~J. {Love}, R.~W. {Heath}, V.~K. {N. Lau}, D.~{Gesbert}, B.~D. {Rao}, and
  M.~{Andrews}, ``An overview of limited feedback in wireless communication
  systems,'' \emph{IEEE Journal on Selected Areas in Communications}, vol.~26,
  no.~8, pp. 1341--1365, 2008.

\bibitem{fpc}
W.~{Xiao}, R.~{Ratasuk}, A.~{Ghosh}, R.~{Love}, Y.~{Sun}, and R.~{Nory},
  ``Uplink power control, interference coordination and resource allocation for
  3gpp e-utra,'' in \emph{IEEE Vehicular Technology Conference}, 2006, pp.
  1--5.

\bibitem{inversion}
H.~{ElSawy} and E.~{Hossain}, ``On stochastic geometry modeling of cellular
  uplink transmission with truncated channel inversion power control,''
  \emph{IEEE Transactions on Wireless Communications}, vol.~13, no.~8, pp.
  4454--4469, 2014.

\bibitem{anum}
A.~{Ali}, N.~{González-Prelcic}, and R.~W. {Heath}, ``Millimeter wave
  beam-selection using out-of-band spatial information,'' \emph{IEEE
  Transactions on Wireless Communications}, vol.~17, no.~2, pp. 1038--1052,
  2018.

\bibitem{doubly}
S.~{Gao}, X.~{Cheng}, and L.~{Yang}, ``Estimating doubly-selective channels for
  hybrid mmwave massive mimo systems: A doubly-sparse approach,'' \emph{IEEE
  Transactions on Wireless Communications}, vol.~19, no.~9, pp. 5703--5715,
  2020.

\bibitem{vq1}
J.~C. {Roh} and B.~D. {Rao}, ``Transmit beamforming in multiple-antenna systems
  with finite rate feedback: a vq-based approach,'' \emph{IEEE Transactions on
  Information Theory}, vol.~52, no.~3, pp. 1101--1112, 2006.

\bibitem{vq2}
W.~{Santipach} and M.~L. {Honig}, ``Achievable rates for mimo fading channels
  with limited feedback,'' in \emph{Eighth IEEE International Symposium on
  Spread Spectrum Techniques and Applications - Programme and Book of Abstracts
  (IEEE Cat. No.04TH8738)}, 2004, pp. 1--6.

\bibitem{adc}
W.~B. {Abbas}, F.~{Gomez-Cuba}, and M.~{Zorzi}, ``Millimeter wave receiver
  efficiency: A comprehensive comparison of beamforming schemes with low
  resolution adcs,'' \emph{IEEE Transactions on Wireless Communications},
  vol.~16, no.~12, pp. 8131--8146, 2017.

\bibitem{adc36}
Y.~{Yu}, P.~G.~M. {Baltus}, A.~{de Graauw}, E.~{van der Heijden}, C.~S.
  {Vaucher}, and A.~H.~M. {van Roermund}, ``A 60 ghz phase shifter integrated
  with lna and pa in 65 nm cmos for phased array systems,'' \emph{IEEE Journal
  of Solid-State Circuits}, vol.~45, no.~9, pp. 1697--1709, 2010.

\bibitem{adc37}
\BIBentryALTinterwordspacing
L.~Kong, ``Energy-efficient 60ghz phased-array design for multi-gb/s
  communication systems,'' Ph.D. dissertation, EECS Department, University of
  California, Berkeley, Dec 2014. [Online]. Available:
  \url{http://www2.eecs.berkeley.edu/Pubs/TechRpts/2014/EECS-2014-191.html}
\BIBentrySTDinterwordspacing

\bibitem{adc38}
{Yu-Hsuan Lin} and H.~{Wang}, ``A low phase and gain error passive phase
  shifter in 90 nm cmos for 60 ghz phase array system application,'' in
  \emph{2016 IEEE MTT-S International Microwave Symposium (IMS)}, 2016, pp.
  1--4.

\bibitem{adc39}
M.~Kraemer, D.~Dragomirescu, and R.~Plana, ``Design of a very low-power,
  low-cost 60 ghz receiver front-end implemented in 65 nm cmos technology,''
  \emph{International Journal of Microwave and Wireless Technologies}, vol.~3,
  no.~2, p. 131–138, 2011.

\bibitem{adc27}
R.~{Méndez-Rial}, C.~{Rusu}, N.~{González-Prelcic}, A.~{Alkhateeb}, and R.~W.
  {Heath}, ``Hybrid mimo architectures for millimeter wave communications:
  Phase shifters or switches?'' \emph{IEEE Access}, vol.~4, pp. 247--267, 2016.

\bibitem{cfo}
R.~W.~H. Jr., \emph{Introduction to Wireless Digital Communication: A Signal
  Processing Perspective}.\hskip 1em plus 0.5em minus 0.4em\relax Prentice
  Hall, 2017.

\bibitem{ni}
\BIBentryALTinterwordspacing
``{mmWave: The Battle of the Bands}.'' [Online]. Available:
  \url{https://www.ni.com/en-us/innovations/white-papers/16/mmwave--the-battle-of-the-bands.html}
\BIBentrySTDinterwordspacing

\bibitem{heath}
R.~W. {Heath}, N.~{González-Prelcic}, S.~{Rangan}, W.~{Roh}, and A.~M.
  {Sayeed}, ``An overview of signal processing techniques for millimeter wave
  mimo systems,'' \emph{IEEE Journal of Selected Topics in Signal Processing},
  vol.~10, no.~3, pp. 436--453, 2016.

\end{thebibliography}

\end{document}